\documentclass[11pt,aps,pra,tightenlines,nofootinbib,longbibliography]{revtex4-2}

\usepackage{amsmath}
\usepackage{amssymb}
\usepackage{bbm}
\usepackage{mathtools}
\usepackage[normalem]{ulem}
\usepackage{dsfont}
\usepackage{mathrsfs}
\usepackage[makeroom]{cancel}
\usepackage{graphicx}
\usepackage{bm}

\DeclareMathOperator\re{Re}
\DeclareMathOperator\im{Im}

\def\x{\boldsymbol{x}}
\def\y{{\boldsymbol{y}}}

\def\rhat{{\hat {\boldsymbol r}}}
\def\khat{{\hat {\boldsymbol k}}}

\def\0{\boldsymbol{0}}

\def\RR{\mathbbm{R}}

\def\J{\boldsymbol{J}}

\def\kk{{\boldsymbol{k}}}
\def\rr{\boldsymbol{r}}

\def\rmJ{{\rm J}}
\def\rmI{{\rm I}}
\def\rmH{{\rm H}}

\def\V{\boldsymbol{V}}
\def\E{\boldsymbol{E}}
\def\EE{\boldsymbol{{\cal E}}}
\def\JJ{\boldsymbol{{\cal J}}}
\def\S{\boldsymbol{S}}
\def\Bold{\boldsymbol{B}}

\def\Lag{{\cal L}}

\def\ii{\mathrm{i}}

\def\dd#1{d^3\mkern-1.5mu#1\,}

\def\bnabla{{\boldsymbol\nabla}}

\def\beps{{\boldsymbol\epsilon}}

\def\vareps#1{\varepsilon_{#1}}
\def\VAREPS#1{\varepsilon_{#1}}

\def\om#1{\omega_{#1}}
\def\OM#1{\omega_{#1}}

\def\smallint{{\textstyle\int}}

\def\la{\langle}
\def\ra{\rangle}

\def\B#1{\left(#1\right)}
\def\BB#1{\left[#1\right]}
\def\BBB#1{\left|#1\right|}

\def\lara#1{\la#1\ra}

\def\be{\begin{equation}}
\def\ee{\end{equation}}

\def\hc{\text{h.c.}}

\def\DIV{\text{div}}

\def\for{\ \text{for} \ }

\def\Xint#1{\mathchoice
{\XXint\displaystyle\textstyle{#1}}%
{\XXint\textstyle\scriptstyle{#1}}%
{\XXint\scriptstyle\scriptscriptstyle{#1}}%
{\XXint\scriptscriptstyle\scriptscriptstyle{#1}}%
\!\int}
\def\XXint#1#2#3{{\setbox0=\hbox{$#1{#2#3}{\int}$}
\vcenter{\hbox{$#2#3$}}\kern-.5\wd0}}

\def\dashint{\Xint-}

\usepackage[utf8]{inputenc}
\usepackage[T1]{fontenc}

\usepackage[ddmmyyyy]{datetime}
\makeatletter
\def\Dated@name{}
\makeatother

\renewcommand{\sinh}{\operatorname{sh}}
\renewcommand{\cosh}{\operatorname{ch}}
\renewcommand{\tanh}{\operatorname{th}}
\renewcommand{\coth}{\operatorname{cth}}

\def\pref{0.575}

\begin{document}

\title{Periodic charge oscillations in the Proca theory}

\author{Bogdan Damski}
\affiliation{Jagiellonian University, 
Faculty of Physics, Astronomy and Applied Computer Science,
{\L}ojasiewicza 11, 30-348 Krak\'ow, Poland}

\begin{abstract}
We consider the Proca theory
of the real massive vector  field. 
There is  a locally conserved $4$-current 
operator
in such a  theory,
which one may  use to define  the charge operator.
Accordingly, there are charged states 
in which the expectation value of 
the charge operator 
is non-zero.
We take a close look at the charge 
operator
and  study the dynamics of the 
certain class of charged states.
For this purpose, we discuss 
the mean electric field and $4$-current 
in such states.
The mean electric field 
has the periodically oscillating 
Coulomb component, whose presence
explains the
periodic charge oscillations.
A complementary insight at such
a phenomenon is provided by the 
mean $4$-current, whose discussion 
leads  to the identification  of 
a certain  paradox.
Last but not  least,  we show that there is  a shock wave                
 propagating in the studied system,
 which affects analyticity of the
mean electric field and $4$-current.
\end{abstract}
\maketitle

\section{Introduction}
\label{Introduction_sec}
The  most economical 
description of the spin-1  vector 
field is delivered by the Proca theory, which 
is best introduced by simply writing the
following
Lagrangian density    \cite{Greiner,Nieto_RMP2010}
\be
\Lag=-\frac{1}{4} 
F_{\mu\nu}F^{\mu\nu}
+\frac{m^2}{2}V_\mu V^\mu,
\label{SPr}
\ee
where 
$F_{\mu\nu}=\partial_\mu V_\nu-\partial_\nu V_\mu$,
$m$ is the mass of the vector 
boson,
and $V$ is the vector field  operator
(we will
be referring to the 
quantum version of the 
Proca theory unless stated otherwise).

Such an exactly solvable
theory--besides delivering
insights into
physics of both 
known particles such as $\rho$ and $\omega$
mesons
and the conjectured ones   such a 
massive photon--provides a fertile playground
for the discussion   of various 
issues   in quantum field theory.
Thus, it should 
come as no surprise 
that it 
appears  in such a context
in 
a dozen or so  textbooks we are aware of  (see e.g. 
\cite{Greiner,ColemanBook,Weinberg}). 
This well-documented 
interest in the Proca theory
makes the impression that all 
of its basic 
aspects 
have   already been 
comprehensively discussed.
This is why we find it 
curious that 
none of the  textbooks  we have 
found 
says  anything about
the charge operator of the
Proca theory and the related issues.

To explain what we are referring  to, 
we write the Proca field equations 
in the following suggestive form
\begin{align}
\label{FJ}
&\partial_\mu F^{\mu\nu}=J^\nu,\\
&J^\nu=-m^2 V^\nu,
\label{Jmu}
\end{align}
from which it is immediately seen that 
the $4$-current operator $J$ satisfies the 
continuity equation
\be
\partial\cdot J=0.
\label{JCont}
\ee
This result leads to the following definition
of the  charge operator
\be
Q(t)=\int \dd{x} J^0(t,\x)=
\int \dd{x} \DIV\E(t,\x),
\label{Q}
\ee
where
\be
\E=\B{F^{i0}}
\label{EFi0}
\ee
 is defined  in the same 
way as the electric field operator in 
the theory of the  massless electromagnetic
field.

The question now is whether the charge 
$\lara{Q}$, 
computed as an expectation
value of operator (\ref{Q}) in a given state, 
is conserved in time. This is a  non-trivial 
question because
(\ref{JCont})  is not expected to guarantee
$\partial_t\lara{Q}=0$ unless $\smallint d\boldsymbol{S}\cdot\lara{\J}$,
the surface integral at 
spatial infinity, is equal to zero (a similar 
requirement  is well known to condition 
charge conservation in other systems
obeying the continuity equation).
The conservation  of $\lara{Q}$ would  clearly
happen  if the charge operator was
manifestly time-independent. However, 
this is not the case because the  charge operator 
in the Proca theory
satisfies the harmonic 
oscillator equation \cite{GHK1964,Guralnik1968_in}.
As a result of that, 
the charge  $\lara{Q}$   is expected to be 
time-independent only  when
it vanishes, which should happen when the mean 
electric field 
exhibits  faster than inverse-square law 
asymptotic decay.
We are interested in this work in a situation,
where the mean electric
field asymptotically follows the inverse-square law
and  the charge   periodically oscillates.
The main stress in our studies  is
placed on (i) the construction of  charged states
in the Proca theory, i.e.,
the states  in which $\lara{Q}\neq0$; (ii) the
discussion of the dynamics of the mean
electric field and $4$-current in such states; 
and (iii) the discussion of periodic charge 
oscillations in the Proca theory from the  perspective of the 
special theory of relativity. 
To the best of our knowledge,
none of the (i)--(iii) issues has been addressed 
in the literature so far (e.g. the harmonic oscillator equation
for $Q(t)$
is stated in 
\cite{GHK1964,Guralnik1968_in} but
no discussion touching upon
the (i)--(iii) points is provided there).


The outline of this paper is the following.
We clarify the charge-related terminology 
in section \ref{Basics_sec}
and summarize 
the conventions that
we use   in appendix 
\ref{Conventions_app}.
We gather the fundamental 
charge operator-related 
results 
 in  section \ref{Dynamics_sec}, where we also
provide some comments placing our research in
a larger context.
We introduce and formally
characterize a certain  class of charged 
states  in
section \ref{Charged_sec}, 
the simplest one illustrating 
the periodic charge oscillations.
We present the thorough discussion of 
the dynamics of one such  state in 
section \ref{Example_sec}.\footnote{While  talking about the 
     dynamics of charged states,
     we tacitly  refer to the 
     Schr\"odinger picture
     formulation of the problem, 
     which is convenient on different occasions. 
     However, our calculations involving 
     field operators
     have been done in the Heisenberg picture,
     and so they are presented in such 
     a standard ``language''.
    }
These considerations are then generalized
in section \ref{Further_sec}, where we 
 comment on the dynamics of the 
whole family of charged states.
The numerous 
technical details, pertinent to 
the  discussion in 
section \ref{Example_sec} 
(section \ref{Further_sec}),
are comprehensively laid out
in appendices \ref{Integral_app} and
\ref{Sign_app} (appendices \ref{Integral_gamma_app}
and \ref{Smoothness_app}).
The conceptual relativity-related issues 
associated with  periodic
charge oscillations 
are discussed in section \ref{Relativity_sec}.
The summary of our work is provided in 
section \ref{Summary_sec}.

\section{Terminology}
\label{Basics_sec}

To begin, we consider the classical 
theory for a moment and mention 
that
the timelike component of 
a locally conserved
$4$-current is traditionally called a
charge
  density in various physical systems.
As a result of that,  
$J^0=\DIV\E=-m^2V^0$ could be termed 
in such a way as well.
A bit different terminology  was proposed 
in \cite{Nieto_RMP2010}, where 
the Proca theory in the presence of 
``ordinary'' charged  particles  was considered.
Then $\DIV\E=-m^2V^0+\rho$
with $\rho$ representing the 
charge density of such particles.
In this  context, 
$-m^2V^0$ was  termed as a pseudocharge 
density.

We shall not dwell on 
which one of these 
two names is
more appropriate.
Coming 
back to the quantum theory, 
we conclude that
$J^0=-m^2V^0$ 
will be    called 
the charge density operator.
In accord with that, $\J=-m^2\V$ 
will be called  the charge current operator,
where  we have 
simplified the terminology 
by skipping the term ``density'' (the same 
simplification has  been applied
in section
\ref{Introduction_sec} to the name 
of the operator $J$). 
$\E$  and $\Bold=\text{rot}\V$ 
will be called 
the electric and magnetic field operators,
without stressing  the fact that they
are defined in the massive theory.
Finally,  for the sake of brevity, 
the expectation 
value of the operators $\E$,  $J$, etc. 
will be called the  mean electric field, 
 the  mean $4$-current, etc.

\section{Charge operator}
\label{Dynamics_sec}
We discuss  here some elementary 
observations
about the charge operator, setting the stage for 
the presentation of our key findings
in the subsequent sections.

To begin, we provide  the expression for the vector field operator
$V$
 \cite{Greiner}
\begin{subequations}
\be
V^\mu(x)= \int
\frac{\dd{k}}{(2\pi)^{3/2}}\frac{1}{\sqrt{2\vareps{k}}}
 \sum_{\sigma=1}^3
\epsilon^\mu(\kk,\sigma) 
[a_{\kk\sigma}\exp(-\ii k\cdot x) + \hc], 
\ee
where $k^0=\vareps{k}=\sqrt{m^2+\om{k}^2}$,  $\om{k}=|\kk|$, 
\be
 [a_{\kk\sigma},a_{\kk'\sigma'}]=0, \
[a_{\kk\sigma},a^\dag_{\kk'\sigma'}]=\delta_{\sigma\sigma'}\delta(\kk-\kk')
\ee
for $\sigma,\sigma'=1,2,3$,
\begin{align}
& \epsilon(\kk,\ell)=(0,\beps(\kk,\ell)), \\ 
& \beps(\kk,\ell)\cdot\kk=0, \ \beps(\kk,\ell)\cdot\beps(\kk,\ell')=\delta_{\ell\ell'}
\end{align}
for $\ell,\ell'=1,2$, and 
\be
\label{eps3}
\epsilon(\kk,3)=\B{\frac{\om{k}}{ m},\frac{\vareps{k}}{m}\khat},
\ \khat=\kk/\om{k}.
\ee
\label{AL}%
\end{subequations}
We mention in passing that the 
transverse polarization
vectors are assumed to be real and 
satisfy 
\be
\sum_{\ell=1}^2
\epsilon^i(\kk,\ell)
\epsilon^j(\kk,\ell)=\delta_{ij}-k^ik^j/\om{k}^2.
\ee

Two operators   can be now
discussed:  the $4$-current operator $J$ (\ref{Jmu})  and the 
 electric
field operator $\E$  \cite{Greiner} 
\begin{align}
\E(x)&=
\ii m\int \frac{\dd{k}}{(2\pi)^{3/2}}
\frac{\khat}{\sqrt{2\vareps{k}}} 
[a_{\kk3} \exp(-\ii k\cdot x) - \hc] \nonumber\\
&+\ii\int \frac{\dd{k}}{(2\pi)^{3/2}}
\sqrt{\frac{\vareps{k}}{2}}\sum_{\sigma=1}^2 
\beps(\kk,\sigma)
[a_{\kk\sigma}\exp(-\ii k\cdot x) - \hc],
\label{piE}%
\end{align}
where  again $k^0=\vareps{k}$.
These operators 
are interlinked via $J^0=\DIV\E$.
Somewhat more interestingly, $J$ and $\E$ are 
also interconnected via 
\be
[E^i(t,\x),J^s(t,\y)]=-\ii  m^2\delta_{is} \delta(\x-\y),
\label{Scw}
\ee
which follows from the canonical commutation
relations 
($E_i$ are the canonical conjugates of the fields $V^i$).
Moreover, it is implied by  (\ref{Scw})
that 
\be
[J^0(t,\x),J^s(t,\y)]=-\ii  m^2
\frac{\partial}{\partial x^s}\delta(\x-\y).
\label{ScwNext}
\ee

Next, we turn 
our attention to the  
charge operator, which 
is obtained by combining  
(\ref{Jmu}), (\ref{Q}), and (\ref{AL}).
The resulting expression, however, 
has to be treated  with some care.
We proceed by considering 
 the following regularized expression
 for the charge operator
\begin{subequations}
\be
\label{1eps1}
\int \dd{x} \exp(-\epsilon|\x|^2)J^0(t,\x)  
=- m(2\pi)^{3/2}
\int\dd{k}\frac{\om{k}\delta_\epsilon(\kk) }{\sqrt{2\vareps{k}}}
[a_{\kk3}\exp(-\ii\vareps{k}t)+\hc],
\ee
where $\epsilon>0$ and 
\be
\delta_\epsilon(\kk)=
\frac{1}{(2\sqrt{\pi\epsilon})^3}
\exp\B{-\frac{\om{k}^2}{4\epsilon}}
\label{nasC}
\ee
\label{qqqQQQ}%
\end{subequations}
is the nascent delta function. 
The above $\epsilon$-regularization
controls the size 
of the spatial integration region,
which is useful  during 
the exchange of the order of spatial ($\dd{x}$) and 
momentum ($\dd{k}$) integrations in (\ref{qqqQQQ}).
The thermodynamic limit 
expression for the charge operator 
is recovered  by
taking $\epsilon$ to $0^+$, which brings us to 
\begin{subequations}
\begin{align}
&Q(t)=\lim_{\epsilon\to0^+} Q_\epsilon(t),\\
&Q_\epsilon(t)= -\sqrt{\frac{ m}{2}}(2\pi)^{3/2}\int\dd{k}
\om{k}\delta_\epsilon(\kk)  [a_{\kk3}\exp(-\ii m t)+\hc].
\label{QQQb}
\end{align}
\label{QQQ}%
\end{subequations}

Explicit time dependence of the
charge operator,
which happens despite the fact that 
the  $4$-current current operator 
satisfies
continuity equation (\ref{JCont}),
provokes the question of what differential 
equation governs the evolution
of $Q(t)$.
It turns out that the charge operator
satisfies the harmonic oscillator 
equation  
\be
\ddot{Q}(t)=- m^2 Q(t),
\label{ddQop}
\ee
where each  dot stands for $d/dt$.
Such a result instantly follows from the 
 differentiation 
of (\ref{QQQ}). 
Alternatively, one may 
obtain it 
by means of the Heisenberg equation 
\begin{subequations}
\begin{align}
&\ddot{Q}(t)=\lim_{\epsilon\to0^+}\BB{H,[Q_\epsilon(t),H]},\\
&H=\int \dd{k} \vareps{k}\sum_{\sigma=1}^3a^\dag_{\kk\sigma}a_{\kk\sigma},
\label{Ham}
\end{align}
\label{ddQNext}%
\end{subequations}
where $H$ is the Hamiltonian of the Proca theory. 
Having said all that, we would like to 
discuss the general context of  results
(\ref{Scw}) and (\ref{ScwNext})--(\ref{ddQop}).

First, (\ref{Scw}) and (\ref{ScwNext})
can be compared to 
the 
Schwinger's prediction about 
the structure of 
electric field--current and 
charge density--current 
commutators \cite{SchwingerPRL1959}.
Namely, based on general considerations,
Schwinger proposed  that
\begin{align}
\label{SCW}
&\lara{[E^i(t,\x),j^s(t,\y)]}=-\ii K^2\delta_{is}\delta(\x-\y),\\
&\lara{[j^0(t,\x),j^s(t,\y)]}=-\ii K^2
\frac{\partial}{\partial x^s}\delta(\x-\y),
\label{SCWNEXT}
\end{align}
where the expectation values are computed 
in the vacuum state,
$K$ is a constant that could be formally infinite, 
and $j$ is the $4$-current in the system
in which  the  
  charge-bearing   matter field, such as 
 the Dirac field 
or the  complex Klein-Gordon  field, 
is coupled to
the electromagnetic field
 \cite{SchwingerPRL1959}.
Note that  (\ref{Scw}) and (\ref{ScwNext}), 
obtained in the theory decoupled from any matter field, 
have exactly the same structure
as (\ref{SCW}) and (\ref{SCWNEXT}) with $K=m$.

Second,
(\ref{qqqQQQ})--(\ref{ddQop})
can be discussed in the context 
of the symmetry
breaking studies 
comprehensively  reviewed in 
\cite{Guralnik1968_in}.
To begin,  
the procedure of 
the evaluation of charge operators 
in the restricted area
of space 
is discussed in \cite{Guralnik1968_in}.
Its particular realization is seen 
in (\ref{qqqQQQ}). 
Next, we note that the
charge operator of the Proca theory
 has 
a singular infrared structure,
just as  charge operators
studied in \cite{Guralnik1968_in}.\footnote{To illustrate  the singular character  of  
     (\ref{QQQ}), we remark that when one tries to compute the norm of the 
     state $Q(t)|0\ra$, one obtains the poorly-defined expression 
     involving the product of two nascent  delta functions 
     having the same momentum argument.
     We note that the problems with the
     computation of such a norm are  expected on general grounds (see 
     \cite{FabriPicasso}  
     as well as  the   comments about this paper
     below equation  (2.5) of  \cite{Guralnik1968_in}).
     }
As it does not lead to  any problems
in our work, we will not dwell on it.
Finally, the harmonic oscillator equation 
for the charge operator of the massive
vector field 
appeared in 
 seminal paper \cite{GHK1964} and 
review \cite{Guralnik1968_in}.
In  these references, however, 
such an  equation 
 is 
merely stated.
Given the popularity
of \cite{GHK1964},
we find it curious that,
to the best of our knowledge, 
the harmonic dynamics of 
the charge in the  Proca theory 
 has not been comprehensively  
 discussed in the literature so far
 (see also the last paragraph of 
 this section 
 for further discussion of this observation).

Till the end of this section,
we  provide  
simple remarks pertaining to  
the harmonic oscillator equation 
satisfied by the charge operator.
The 
general solution of  
(\ref{ddQop}) reads  
\be
Q(t)=Q(0)\cos( m t) + \frac{\dot{Q}(0)}{ m}\sin( m t),
\label{QcosinOP}
\ee
from which the  periodic charge conservation law
is instantly obtained. Defining 
\be
{\cal Q}(t)=\la Q(t)\ra, \ \dot{{\cal Q}}(t)=\la \dot{Q}(t)\ra,
\ee
where the expectation value is taken in a
normalizable time-independent
Heisenberg-picture state, we get
\be
{\cal Q}(t)={\cal Q}(0)\cos( m t) + \frac{\dot{\cal Q}(0)}{ m }\sin( m t).
\label{Qcosin}
\ee

The usual charge conservation law, $\dot{{\cal Q}}(t)=0$, is only 
seen when 
\be
{\cal Q}(0)=\dot{\cal Q}(0)=0.
\label{QdQ}
\ee
For example, this 
is the case in    any state of the form
\be
c|0\ra + \int\dd{k}\sum_{\sigma=1}^2 c_\sigma(\kk)a^\dag_{\kk\sigma}|0\ra
 +\int\dd{k}\dd{k'}\sum_{\sigma,\sigma'=1}^2 c_{\sigma\sigma'}(\kk,\kk')
a^\dag_{\kk\sigma}
a^\dag_{\kk'\sigma'}|0\ra+ \cdots
\ee
or 
\be
\int \dd{k} \sum_{\sigma=1}^3 c_\sigma(\kk) a^\dag_{\kk\sigma}|0\ra, 
\ee 
where $|0\ra$ is the vacuum state of the Proca theory and 
the complex-valued 
$c$, $c_\sigma(\kk)$, $c_{\sigma\sigma'}(\kk,\kk')$, etc. 
are constrained  so as to make the above states normalizable.

A markedly different situation takes place 
when  (\ref{QdQ}) is not satisfied. 
The total charge is only 
periodically conserved then.
Its oscillation period,    
\be
\frac{2\pi}{m},
\label{PerioD}
\ee
depends  on the mass of the vector boson  
described by the Proca
theory. We  note  that a similar scenario
was  mentioned in \cite{Hertzberg2020} in the context 
of  spatially unbounded superconductors.
In these non-relativistic mean-field  considerations,
the role of the mass $m$ is played by the 
plasma frequency(ies).

We are interested in this work in finding 
the detailed explanation of 
the periodic charge oscillations
in the Proca theory.
For that purpose, we will 
analyze the dynamics of the states 
in which the 
long-distance decay of the 
mean electric field  follows the 
inverse-square law. To the best of our knowledge,
such studies  have not 
been reported  before. This observation may be 
explained by the fact that the short-range
electric field is traditionally
discussed in the context of the massive photon 
theories.
It is presumably so because   localized external sources
are expected in such theories  to create the
exponentially decaying electric field 
(the total charge for such a field vanishes 
and as such it  is conserved;
see   \cite{Hertzberg2019} for a concrete example
where not only the total charge but also the one enclosed in a 
finite volume is discussed in the presence of the short-range 
electric field).
However, it should be stressed that  
well-defined finite-energy
normalizable 
states, in which the mean electric field 
   asymptotically  satisfies   
the
inverse-square law,
{\it do} exist in the Hilbert space of the Proca theory.
As the periodic charge conservation law implies, they
cannot be stationary. 
The theoretical characterization of their 
non-trivial 
dynamics is the main goal of this work.

\section{Charged  states}
\label{Charged_sec}

Charged  states in the Proca theory 
can be  constructed 
in  an innumerable number of ways.
The simplest of them are 
built of   the vacuum state 
and the Fock states containing 
just one  longitudinal field 
excitation. The particular 
  class of such states 
will be commented upon below.

To proceed, we introduce 
\be
\chi(\x)=
\frac{\ii e}{m} \int\frac{\dd{k}}{(2\pi)^{3/2}}
f(\OM{k})
\sqrt{\frac{\VAREPS{k}}{2\OM{k}^2}}
(a_{\kk3}\exp(\ii\kk\cdot\x) - \hc),
\label{tem}
\ee
where $e$ is the electric charge unit and 
$f$ is a dimensionless real function.
The form of $\chi$ 
 is chosen such that
\be
\BB{  \int \dd{y} \exp\B{-\epsilon|\y|^2}J^0(t,\y), \chi(\x)   }
=\ii e
\int\dd{k}f(\OM{k})\delta_\epsilon(\kk)\cos(\vareps{k}t+\kk\cdot\x),
\label{Qetem}
\ee
which guarantees 
\be
[Q(t),\chi(\x)]=\ii e f(0)\cos(mt).
\label{Qcomu}
\ee
This suggests 
consideration of the state  
\be
\label{hpsi}
 |\psi(\x)\ra=\BB{\alpha -\ii  \beta \chi(\x)}|0\ra,\\
\ee
where the 
dimensionless real parameters   $\alpha,\beta$ are nonzero.
Two  remarks are in order now.

First,  we require that   $\la\psi|\psi\ra=1$,
which   leads to  
\be
\label{nhpsi}
1=\alpha^2+\beta^2\frac{e^2}{4\pi^2m^2}
\int_0^\infty d\OM{k}f^2(\OM{k})\VAREPS{k}.
\ee
Such a relation is meaningful as long
as the above integral  is convergent.
This 
 will be guaranteed by 
 asking that 
 the mean energy is finite
\be
\label{Hhpsi}
{\cal H}=\la\psi(\x)|H|\psi(\x)\ra=\beta^2\frac{e^2}{4\pi^2 m^2}
\int_0^\infty d\OM{k}f^2(\OM{k})\VAREPS{k}^2<\infty.
\ee

Second, the 
mean  charge stored
in state (\ref{hpsi}) can be easily computed 
by means of (\ref{Qcomu})  
\be
{\cal Q}(t)=
\la\psi(\x)|Q(t)|\psi(\x)\ra=
\alpha\beta f(0) e \cos(mt).
\label{Qttq}
\ee
This  expression 
underscores  the role of the  IR sector in  the
discussed problem and it tells us that
$0<|f(0)|<\infty$ is of interest
in the context of 
periodic charge oscillations.
We
employ the following  ``normalization'' of the function 
$f$ 
\be
f(0)=1.
\label{f01}
\ee
Upon such a choice, the amplitude of periodic charge oscillations 
reads 
\be
q=\alpha\beta e.
\label{qabe}
\ee
Further insights into the periodic charge oscillations 
can be obtained from two interrelated quantities:
the mean electric field and
  $4$-current.

The mean electric field--obtained by combining (\ref{piE}), (\ref{tem}), 
(\ref{hpsi}), and (\ref{qabe})--reads 
\be
\label{Ehpsi}
\EE(t,\rr)=\la\psi(\x)|\E(t,\y)|\psi(\x)\ra=
-\frac{q\rhat}{2\pi^2}\int_0^\infty
d\OM{k} f(\OM{k})\partial_r\frac{\sin(\OM{k}r)}{\OM{k}r}\cos(\VAREPS{k}t),
\ee
where $\rhat=\rr/|\rr|$, $r=|\rr|$,  and  $\rr=\y-\x$. 
This expression,
 under the assumption that the derivative can be taken 
outside the integral\footnote{Such an operation can be performed without
     any complications in 
     specific yet fairly representative
     problems discussed
     in sections \ref{Example_sec}
     and \ref{Further_sec}.
     We  shall not dwell on
     the general conditions
     allowing for it.},  
can be cast into the following 
suggestive  form
\begin{subequations}
\begin{align}
\label{EhpsiNew}
&\EE(t,\rr)=-\bnabla\phi(t,r),\\
&\phi(t,r)=\frac{q}{2\pi^2r}\int_0^\infty
d\OM{k} f(\OM{k})\frac{\sin(\OM{k}r)}{\OM{k}}\cos(\VAREPS{k}t).
\label{hatphi}
\end{align}
\label{calE}%
\end{subequations}

The mean $4$-current reads
\begin{align}
\label{calJ0}
&{\cal J  }^0(t,r) = \la\psi(\x)|-m^2V^0(t,\y)|\psi(\x)\ra = \DIV\EE(t,\rr), \\
&\JJ(t,\rr)=\la\psi(\x)|-m^2\V(t,\y)|\psi(\x)\ra=-\partial_t{\EE}(t,\rr),
\label{calJ}
\end{align}
where one has to be aware that an
ambiguity may arise 
during the evaluation of (\ref{calJ0}) and
(\ref{calJ}); see section \ref{Shock_sub} 
for details.

Non-trivial  dynamics captured by
(\ref{calE})--(\ref{calJ}) 
will be comprehensively 
analyzed  in a specific 
case in the next section.
Before jumping right there, however, we would like to briefly
comment on the following issues.

First, $\phi$ plays the role of the electric field potential
in (\ref{calE}). 
Second, the last equality in (\ref{calJ0}) 
originates from  $J^0=\DIV\E$.
Third,  the  last equality in (\ref{calJ})
can be explained by combining the 
Proca equation
$\text{rot}\Bold=\J+ \partial_t\E$ 
with the observation that 
the mean magnetic field  
trivially vanishes 
in the studied  states because 
no transverse modes are populated
in them.

\section{Example of periodic
charge oscillations}
\label{Example_sec}

We  illustrate here the concepts 
introduced in section \ref{Charged_sec}.
For this purpose, we  
consider  
\be
f(\OM{k})=\B{\frac{ m}{\VAREPS{k}}}^2,
\label{hatfL2}
\ee
which   leads to 
an elegant analytical 
solution.

To begin, we mention that the evaluation of 
(\ref{nhpsi}) and (\ref{Hhpsi}) yields
\begin{align}
\label{alfaalfa}
&1=\alpha^2 + \beta^2 \frac{e^2}{4\pi^2}, \\
&{\cal H}= \beta^2  m\frac{e^2}{8\pi}.
\label{Hg2}
\end{align}
The integrals leading to such results are 
straightforwardly computed 
with the help of  
 mapping (\ref{mappingA}) and the 
 following formula
\be
\int_0^\infty \frac{dx}{\cosh^b(x)}=\frac{\sqrt{\pi}}{2}
\frac{\Gamma(b/2)}{\Gamma(b/2+1/2)} \for b>0,
\label{chb}
\ee
which can be obtained  from  expression  3.518.3 listed in  \cite{Ryzhik}.

Combining  (\ref{alfaalfa})  with (\ref{qabe}), we find that 
\begin{align}
\label{aa}
&\alpha=\frac{\exp(\ii\sigma)}{\sqrt{2}}\sqrt{1\pm\sqrt{1-\frac{q^2}{\pi^2}}}, \\
&\beta=\sqrt{2}\frac{\pi}{e}\sqrt{1\mp\sqrt{1-\frac{q^2}{\pi^2}}},
\label{bb}
\end{align}
where $\sigma$ is equal to $0$ ($\pi$) for $q$ greater (smaller) than zero.
These expressions  
show that the amplitude of periodic charge oscillations
in the studied problem 
is upper bounded by $\pi$. There is 
nothing fundamental about such an 
elegant bound because it 
depends on the particular form of the function $f$
(section \ref{Further_sec}).

Next, we note that the electric field potential in the 
studied problem reads
\be
\phi(t,r)=\frac{q}{2\pi^2r}\int_0^\infty
d\OM{k} \B{\frac{ m}{\VAREPS{k}}}^2 \frac{\sin(\OM{k}r)}{\OM{k}}\cos(\VAREPS{k}t).
\label{hatphiL2}
\ee
The hint of what we are dealing here with
comes from
\be
\phi(0,r)=q \frac{1-\exp(- m r)}{4\pi r},
\label{phi0r}
\ee
which has been obtained  from (\ref{IabZERO}).
This result  implies   that initially, i.e. at $t=0$, 
there is  the  Coulomb
field in our system from which the Debye  field
is subtracted. The former field certainly
needs no introduction,
the latter  typically appears  
in the studies of plasmas and electrolytes, where 
it captures the screening of the Coulomb field of the
test charge inside such systems.
Equation (\ref{phi0r}) triggers two remarks.

Technically, the above-mentioned  subtraction
makes the energy finite and it can be seen as one of the many examples
of how the Coulomb field may be regularized at short distances.
Note that while other regularizations  can be introduced by changing 
the function $f$, the Coulomb component
of (\ref{phi0r}) is fixed by the requirement that
${\cal Q}(0)=q$ [see (\ref{Qqcos})].

Physically, neither the Coulomb nor  Debye  field is external
source-generated in the studied problem. This implies 
that 
the mean electric field will necessarily evolve in time. 

Having discussed the situation at $t=0$, we are ready to 
study the dynamics for $t>0$.
The evaluation of the integral from (\ref{hatphiL2}) is presented 
in appendix \ref{Integral_app}. It leads to the following result. 
For $r\ge t>0$
\be
\phi(t,r)= q\frac{\cos( m t)-\exp(- m r)}{4\pi r},
\label{phi_large}
\ee
whereas for  $0<r<t$
\be
\phi(t,r)=q\frac{\cos(mr)-\exp(-mr)}{4\pi r} 
-\frac{q}{4\pi r}
\int_{mr}^{mt} dy 
\int_0^{mr} dx
\rmJ_0\B{ \sqrt{y^2-x^2}},
\label{phi_small}
\ee
where 
$\rmJ_n$ is the Bessel function of the first kind
of order $n$
(it should
not be confused with the $n$-th
covariant component of the $4$-current
operator $J$).
The following conclusions follow from the analysis of these
expressions.

\subsection{Shock wave front}
\label{Shock_sub}

The curious division  of  space into the 
$r\ge t$ and  $0<r<t$
regions,  exhibited by (\ref{phi_large}) and (\ref{phi_small}),
comes from the fact that $\phi(t,r)$ is singular 
on the sphere $r=t$. It turns out that 
there is a 
 shock wave in $\phi(t,r)$  propagating 
outwards from the point $\rr=\0$ with the speed 
of light. 
Namely, while   
 $\phi(t,r)$, $\partial_r\phi(t,r)$,
 and $\partial_t\phi(t,r)$
 are continuous across $r=t$,
 the second order derivatives  of  $\phi(t,r)$
 are discontinuous there. 
The subtle issues associated with the
differentiation of $\phi(t,r)$ 
are also 
seen from the fact that while 
\be
\partial_\xi
\int_0^\infty
d\OM{k} \B{\frac{ m}{\VAREPS{k}}}^2
\frac{\sin(\OM{k}r)}{\OM{k}}\cos(\VAREPS{k}t)
 =  \int_0^\infty
d\OM{k} \B{\frac{ m}{\VAREPS{k}}}^2 
\partial_\xi \B{
\frac{\sin(\OM{k}r)}{\OM{k}}\cos(\VAREPS{k}t)}
\label{dal}
\ee
holds for any $r,t>0$, 
\be
\partial_{\eta}\partial_\xi
\int_0^\infty
d\OM{k} \B{\frac{ m}{\VAREPS{k}}}^2
\frac{\sin(\OM{k}r)}{\OM{k}}\cos(\VAREPS{k}t)
 =  \int_0^\infty
d\OM{k} \B{\frac{ m}{\VAREPS{k}}}^2 
\partial_{\eta}\partial_\xi
\B{
\frac{\sin(\OM{k}r)}{\OM{k}}\cos(\VAREPS{k}t)
}
\label{dalbe}
\ee
holds for $r,t>0$  expect   $r=t$ 
($\xi$ represents  either $t$ or $r$ and so does $\eta$).
 These  observations 
 come  from the results presented in 
appendices \ref{Disc_app} and \ref{Sign_app}.
The following remarks are in order now.

First, 
(\ref{dal}) guarantees that $\lara{\psi|\E|\psi}=-\bnabla\phi$ 
for all $r,t>0$. Such a mean electric 
field--with the help of (\ref{phi_large}), (\ref{phi_small}), 
and  the results presented in appendix 
\ref{First_app}--can be written in the following
suggestive form 
\be
\EE(t,\rr)=\rhat{\cal A}(t,r)+\frac{qm^2\rhat}{8\pi t}|t-r| 
+O\B{(t-r)^2},
\label{Eabs}
\ee
where ${\cal A}(t,r)$,
$\partial_t{\cal A}(t,r)$, and $\partial_r{\cal A}(t,r)$
are continuous in the neighborhood
of $r=t$.

Second,  we turn our attention to the mean $4$-current.
The equality in (\ref{dalbe}) for $r\neq t$ 
means that for such $r,t$   we have 
$\protect{{\cal J}^0=\lara{\psi|-m^2V^0|\psi}=\DIV\EE}$
and $\JJ=\lara{\psi|-m^2\V|\psi}=  -\partial_t\EE$.
A quick look at
(\ref{Eabs}) reveals that these two quantities 
must be discontinuous across $r=t$. 
Such a  propagating  discontinuity, 
i.e. a shock wave, is 
illustrated by the following formulas
\begin{align}
\label{sHH1}
&\frac{qm^2}{4\pi t}=
\lim_{r\to t^+}{\cal J}^0(t,r)    -\lim_{r\to t^-}{\cal J}^0(t,r),\\
&\frac{qm^2\rhat}{4\pi t}=
\lim_{r\to t^+}\JJ(t,r\rhat)  -\lim_{r\to t^-}\JJ(t,r\rhat),
\label{sHH2}
\end{align}
which   can be derived  from 
  (\ref{Eabs}).

Third, exactly at the $r=t$ shock wave front, 
we find with the help of the results presented in
appendix \ref{Sign_app2}   that 
$\lara{\psi|-m^2V^0|\psi}$ and 
$\lara{\psi|-m^2\V|\psi}$ are given by 
\be
\label{J0Jszok}
q m^2  \frac{\exp(-m t)-1/2}{4\pi t}, 
\  
q m\rhat \frac{\sin(mt)-mt/2}{4\pi t^2}.
\ee
At the same time, however, 
\be
\DIV\EE=\text{undefined}, \  -\partial_t\EE= \text{undefined}
\label{J0Jmean}
\ee
at $r=t$, which is   seen from (\ref{Eabs}).
These results lead to the
following observations.

To begin, we note that 
the discrepancy between (\ref{J0Jszok}) 
and (\ref{J0Jmean}) 
implies that the last equalities 
in (\ref{calJ0}) and (\ref{calJ})
do not hold at $r=t$ when $f(\OM{k})$
is given by
(\ref{hatfL2}). This  is caused by 
the fact that 
(\ref{dalbe}) does not hold 
at $r=t$. 

Then, we remark that the difference between 
(\ref{J0Jszok})
and (\ref{J0Jmean})
raises the question of whether  
${\cal J}^0(t,t)$ and $\JJ(t,t\rhat)$
can be unambiguously computed.
As we are unsure how to 
convincingly address this issue, we
shall leave open the question of 
what is the value of 
${\cal J}^0(t,t)$ 
and $\JJ(t,t\rhat)$ in the 
considered problem.
We only mention that the discussed 
ambiguity is  not generic in the sense
that it  can be removed by increasing 
the decay rate  of $f(\om{k}\to\infty)$.
This is illustrated in section \ref{Further_sec},
where ${\cal J}^0(t,t)$
and $\JJ(t,t\rhat)$ are straightforwardly
computed.

All in all, we find it remarkable that there is a genuine 
shock wave discontinuity 
in the studied problem 
despite the fact that
the energy associated with 
the analyzed solution is finite (\ref{Hg2}).

\subsection{From shock wave front to spatial infinity}
\label{From_sub}
We are interested here in the region of space $r > t$, 
where (\ref{phi_large}) holds. The
mean electric field obtained from such an 
expression reads 
\be
\EE(t,\rr)=\frac{q\rhat}{4\pi r^2}\cos(mt) -
\frac{q\rhat \exp(-mr)}{4\pi r^2}(1+mr).
\label{Elarge}
\ee
Several remarks are in order now.

First of all, the first term in (\ref{Elarge}) 
represents the 
{\it periodically oscillating} Coulomb field.
Such a field, to the best of our knowledge, has
not been previously discussed in the literature.
The second term in (\ref{Elarge})  
is the Debye field, which 
is frozen in its $t=0$ arrangement
(see section \ref{Further_sec} for the 
dynamical but aperiodic  version of such 
a field).

Second, the Debye component 
of (\ref{Elarge}) 
is short-range and so it does not affect the 
periodic charge oscillations. 
Their amplitude and 
period  can be instantly 
read  from 
the Coulomb component of (\ref{Elarge})
because 
\be
\label{Qqcos}
{\cal Q}(t)=\lim_{r\to\infty}\int d\S(\rr)\cdot 
\EE(t,\rr)  = q\cos( m t),
\ee 
where $d\S$ is the surface element.
It should be now stressed that such 
a result has been obtained 
without the use of any  regularization. 
Therefore, it supports the
$\epsilon$-regularization
procedure discussed in 
section \ref{Dynamics_sec} 
because the very same result 
is also obtained by computing the 
expectation value of (\ref{QQQ}).

\begin{figure}[t]
\centering
\includegraphics[width=\pref\columnwidth,clip=true]{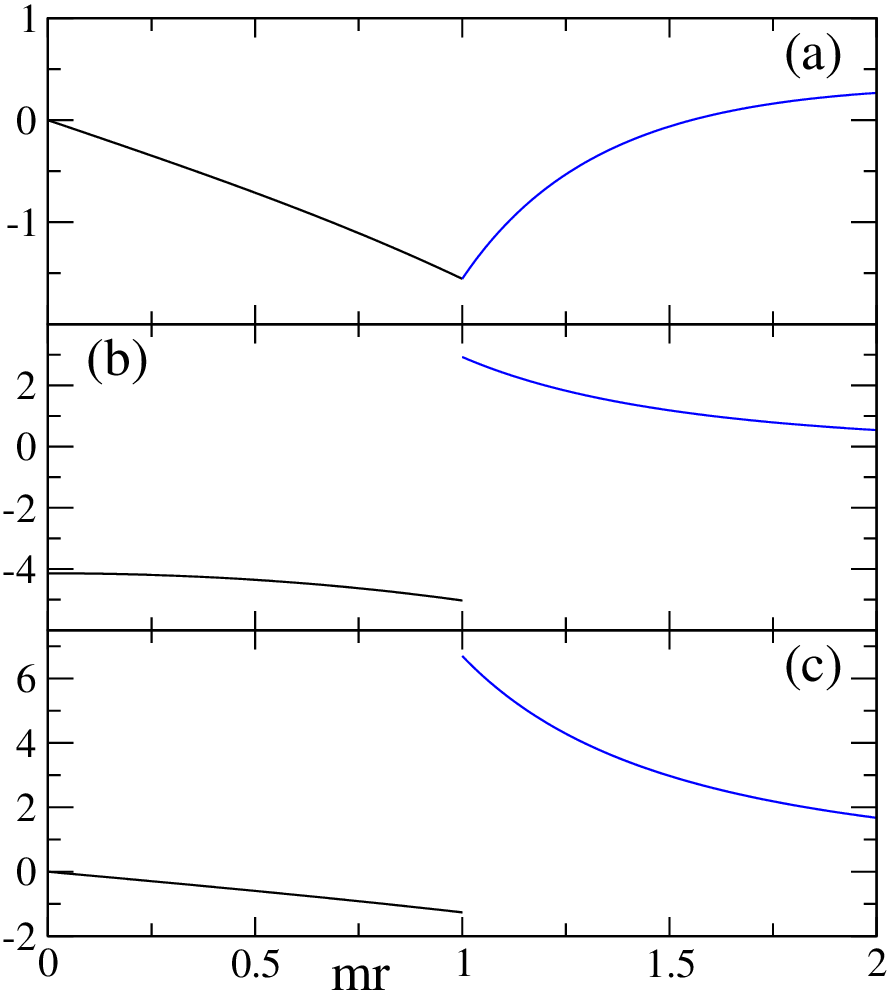}
\caption{The mean values of various quantities
characterizing the properties of the system
at  $m t=1$.
Panel (a):   the rescaled radial component of
 the electric field,
$\EE\cdot\rhat\times 10^2 q^{-1}m^{-2}$
obtained from (\ref{Elarge}) and (\ref{Ebehind}).
Panel (b):
the rescaled  charge density, 
${\cal J}^0\times 10^2 q^{-1}m^{-3}$
obtained from (\ref{J0far}) and 
(\ref{J0behind}).
Panel (c): 
the rescaled radial component 
of the charge current, 
$\JJ\cdot\rhat\times 10^2 q^{-1}m^{-3}$
obtained from (\ref{Jfar}) and 
(\ref{Jvecbehind}).
The  black and  blue lines show $r<t$ and $r>t$ results,
respectively. 
}
 \label{EJ0Jvec_fig}
\end{figure}

Third, the charge stored 
in the $r > t$ region of space  is 
\be
{\cal Q}(t,r > t)= \int_{r>t}\dd{r}{\cal J}^0(t,r)=
q \exp(-mt)(1+mt).
\label{Qfar}
\ee
It becomes exponentially small 
for $mt\gg1$.

Fourth, the mean $4$-current  reads 
\begin{align}
\label{J0far}
&{\cal J}^0(t,r)=\frac{q m^2}{4\pi r}\exp(-m r),\\
&\JJ(t,\rr)=\frac{qm\rhat}{4\pi r^2} \sin( m t).
\label{Jfar}
\end{align}
A trivial calculation confirms that $\partial\cdot{\cal J}=0$.

\begin{figure}[t]
\centering 
\includegraphics[width=\pref\columnwidth,clip=true]{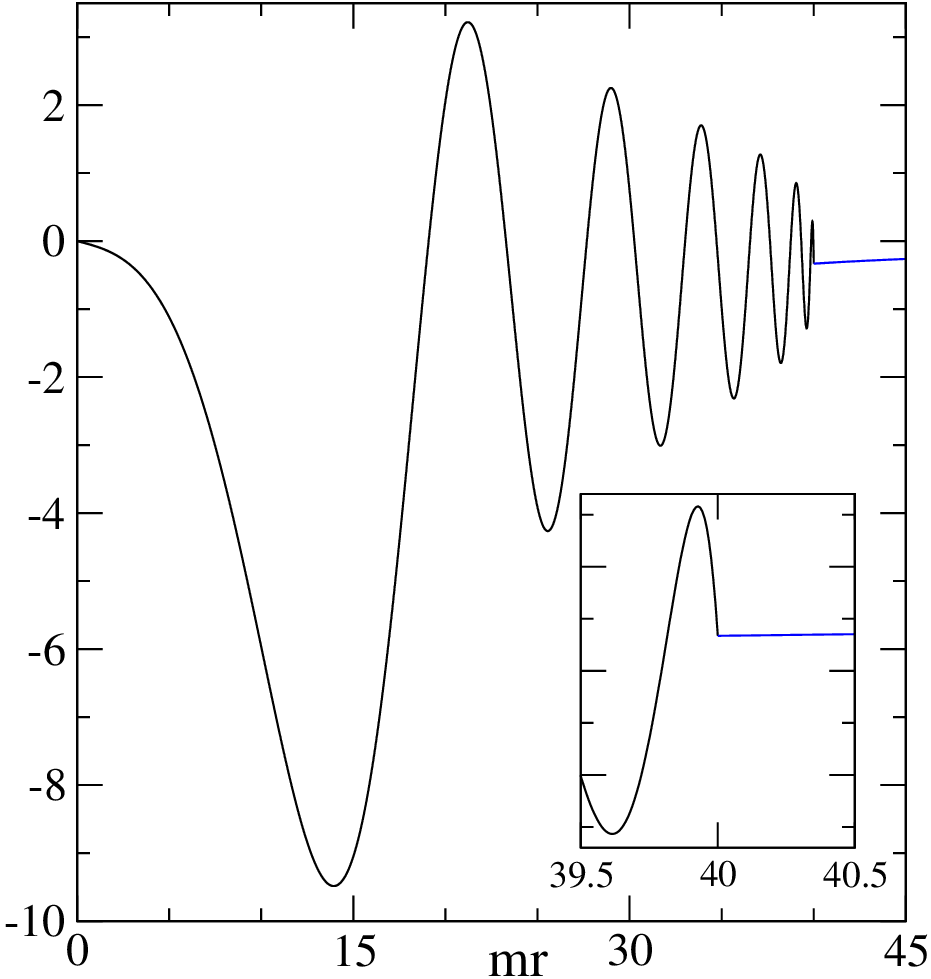}
\caption{The rescaled radial component of
the mean electric field.
Namely, $\EE\cdot\rhat\times 10^4 q^{-1}m^{-2}$ at 
$m t=40$.
The  black line comes from
(\ref{Ebehind}),
whereas the blue one is obtained  from  (\ref{Elarge}).
The inset magnifies the area around the shock
wave front, where the slope of the plotted quantity 
discontinuously changes.
}
 \label{E_fig}
\end{figure}

\subsection{Behind shock wave front}
\label{Behind_sub}

The $0< r<t$ region of space is of interest here, where 
the mean electric field and $4$-current 
are   given by the following formulas
\begin{align}
\label{Ebehind}
&\EE(t,\rr)=\frac{\rhat\phi(t,r)}{r}-\frac{q m \rhat\exp(-mr)}{4\pi r}
+\frac{qm\rhat}{4\pi r}\int_{mr}^{mt} dx \rmJ_0\B{\sqrt{x^2-(mr)^2}},\\
&{\cal J}^0(t,r)=\frac{q m^2}{4\pi r}\exp(-m r) - \frac{q m^2}{4\pi r}
 +\frac{q m^3}{4\pi}\int^{mt}_{mr} dx\frac{\rmJ_1\B{\sqrt{x^2-(mr)^2}}}{\sqrt{x^2-(mr)^2}},
\label{J0behind}\\
&\JJ(t,\rr)=\frac{q m \rhat}{4\pi r^2} \int_0^{mr} dx \rmJ_0\B{\sqrt{(mt)^2-x^2}}
- \frac{q m^2 \rhat}{4\pi r} \rmJ_0\B{m\sqrt{t^2-r^2}}.
\label{Jvecbehind}
\end{align}
Several remarks are in order now. 

First, these results are obtained 
from electric field
potential (\ref{phi_large}),
which explicitly appears in (\ref{Ebehind}). 
As can be  checked,   
(\ref{J0behind}) and (\ref{Jvecbehind}) satisfy
$\partial\cdot{\cal J}=0$.

\begin{figure}[t]
\centering 
\includegraphics[width=\pref\columnwidth,clip=true]{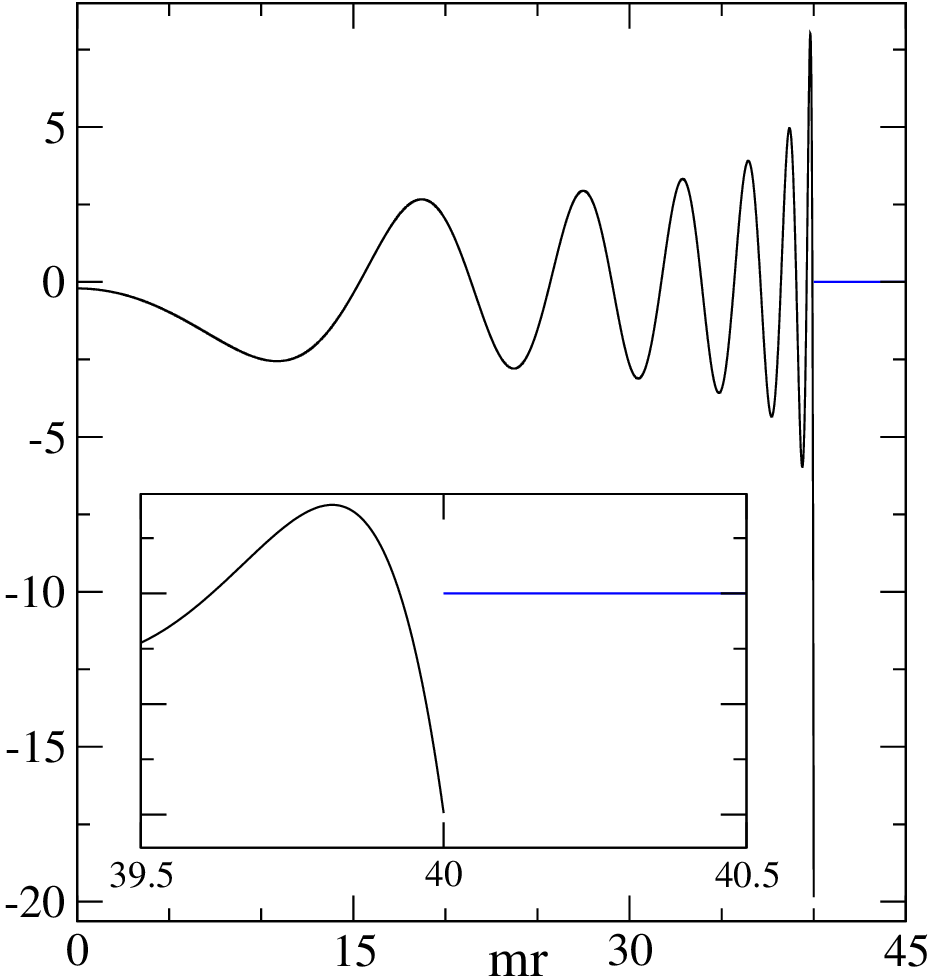}
\caption{The rescaled mean charge density.
Namely, ${\cal J}^0\times 10^4 q^{-1}m^{-3}$ at
$m t=40$.
The  black line comes from
(\ref{J0behind}),
whereas the blue one is obtained  from  (\ref{J0far}).
The inset magnifies  the shock wave discontinuity.
}
 \label{J0_fig}
\end{figure}

\begin{figure}[t]
\centering 
\includegraphics[width=\pref\columnwidth,clip=true]{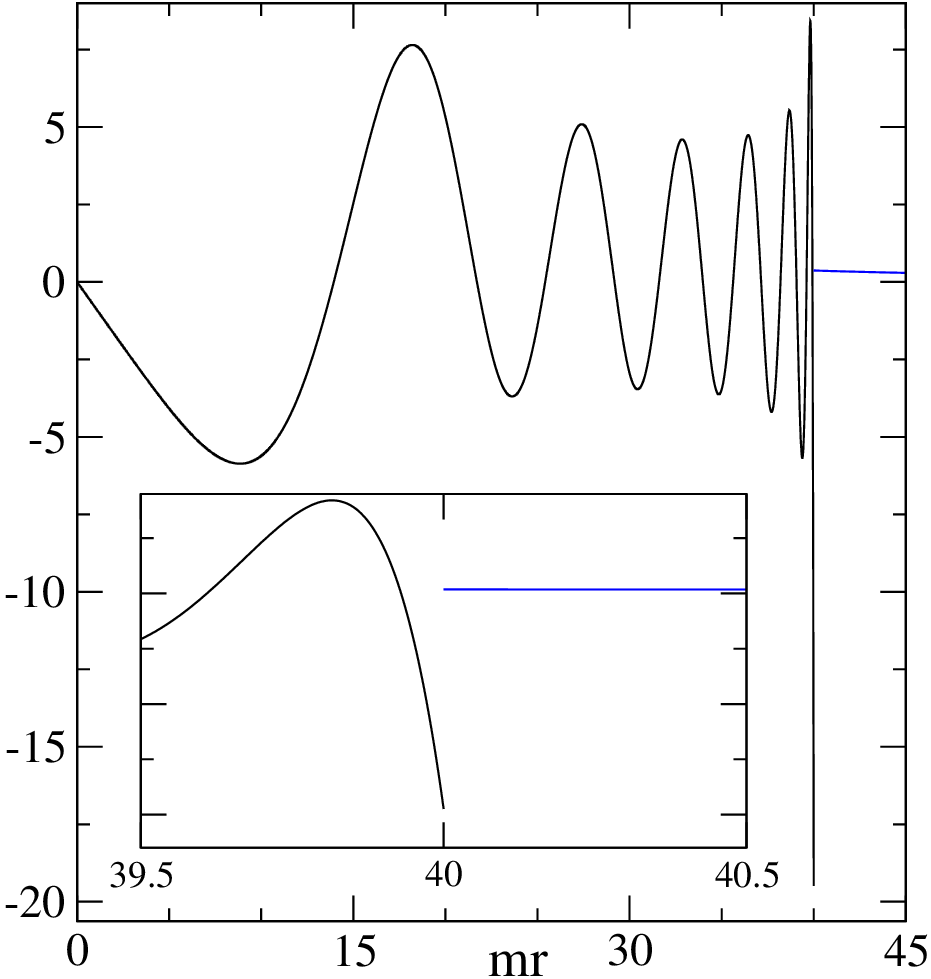}
\caption{The rescaled radial component of 
the mean charge  current.
Namely, $\JJ\cdot\rhat\times 10^4 q^{-1}m^{-3}$ at 
$m t=40$.
The  black line comes from
(\ref{Jvecbehind}),
whereas the blue one is obtained  from  (\ref{Jfar}).
The inset magnifies  the shock wave discontinuity.
}
 \label{Jvec_fig}
\end{figure}

Second, as these quantities are given by fairly 
complicated expressions, we plot them 
on figures \ref{EJ0Jvec_fig}--\ref{Jvec_fig} 
(note that the rescaling factors
used in  figure \ref{EJ0Jvec_fig} are 
two orders of magnitude smaller 
than the ones employed in 
figures \ref{E_fig}--\ref{Jvec_fig}).
Figure \ref{EJ0Jvec_fig}  provides the
representative 
illustration of  (\ref{Ebehind})--(\ref{Jvecbehind})
in the short-time limit. Figures \ref{E_fig}--\ref{Jvec_fig}
 representatively
display what happens  in the long-time limit.
Namely,  the shock wave front leaves
behind spatial oscillations 
of $\EE$, ${\cal J}^0$,
and $\JJ$.
As can be numerically verified, 
the number of such oscillations 
increases as the time goes by while 
their amplitude
decreases.

Third, the above-mentioned oscillations are 
quite dramatic.
Indeed,  for large-enough times 
the mean electric field and charge current  
reverse their direction several times in the 
studied region of space. 
Similarly, the charge density  
changes its sign 
 several times there.
Clearly, a curious dynamics
is seen in figures \ref{E_fig}--\ref{Jvec_fig}, 
 strikingly different from the one 
 discussed in section \ref{From_sub}.

Formulas  (\ref{Ebehind})--(\ref{Jvecbehind})
 can be simplified
near $r=0$ and $r=t^-$.
In the latter case 
\begin{align}
\label{Econti}
&\lim_{r\to t^-}\EE(t,r\rhat)=q\rhat\frac{\cos(mt)-(1+mt)\exp(-mt)}{4\pi t^2},\\
\label{J0shock}
&\lim_{r\to t^-}{\cal J}^0(t,r)=q m^2\frac{\exp(-m t)-1 }{4\pi t},\\
&\lim_{r\to t^-}\JJ(t,r\rhat)=q m\rhat \frac{\sin(mt)-mt}{4\pi t^2}.
\label{Jvecshock}%
\end{align}
The first two of these expressions 
can be easily  read off
(\ref{Ebehind}) and (\ref{J0behind}) while the 
last one has been obtained from (\ref{Jvecbehind})
with the help of 
\be
\int_0^{mt} dx \rmJ_0\B{\sqrt{(mt)^2-x^2}} = \sin(mt),
\label{J0sin}
\ee
see formula 6.517 listed in
\cite{Ryzhik}.

Near $r=0$ the following expressions
describe the quantities of interest
\begin{align}
\label{Esm}
&\EE(t,\rr)=-\frac{qm^3\rhat}{12\pi}F(mt)r +O\B{r^3},\\
\label{J0sm}
&{\cal J}^0(t,r)=-\frac{q m^3}{4\pi } F(mt) + O\B{r^2},\\
&\JJ(t,\rr)=-\frac{q m^4 \rhat}{24\pi}\BB{\rmJ_0(mt)+\rmJ_2(mt)}r 
+ O\B{r^3},
\label{Jvecsm}
\end{align}
where
\be
\label{Ffunc}
F(x)= 1 + \rmJ_1(x) -x \rmJ_0(x)
-\frac{\pi x}{2}\rmJ_1(x)\rmH_0(x)
+\frac{\pi x}{2}\rmJ_0(x)\rmH_1(x)
\ee
with $\rmH_n$  representing the Struve function 
of order $n$
(the last three
terms  
of (\ref{Ffunc}) correspond to  $-\smallint^x_0 dx \rmJ_0(x)$;
see expression 6.511.6 listed in
\cite{Ryzhik}). 

\begin{figure}[t]
\centering 
\includegraphics[width=\pref\columnwidth,clip=true]{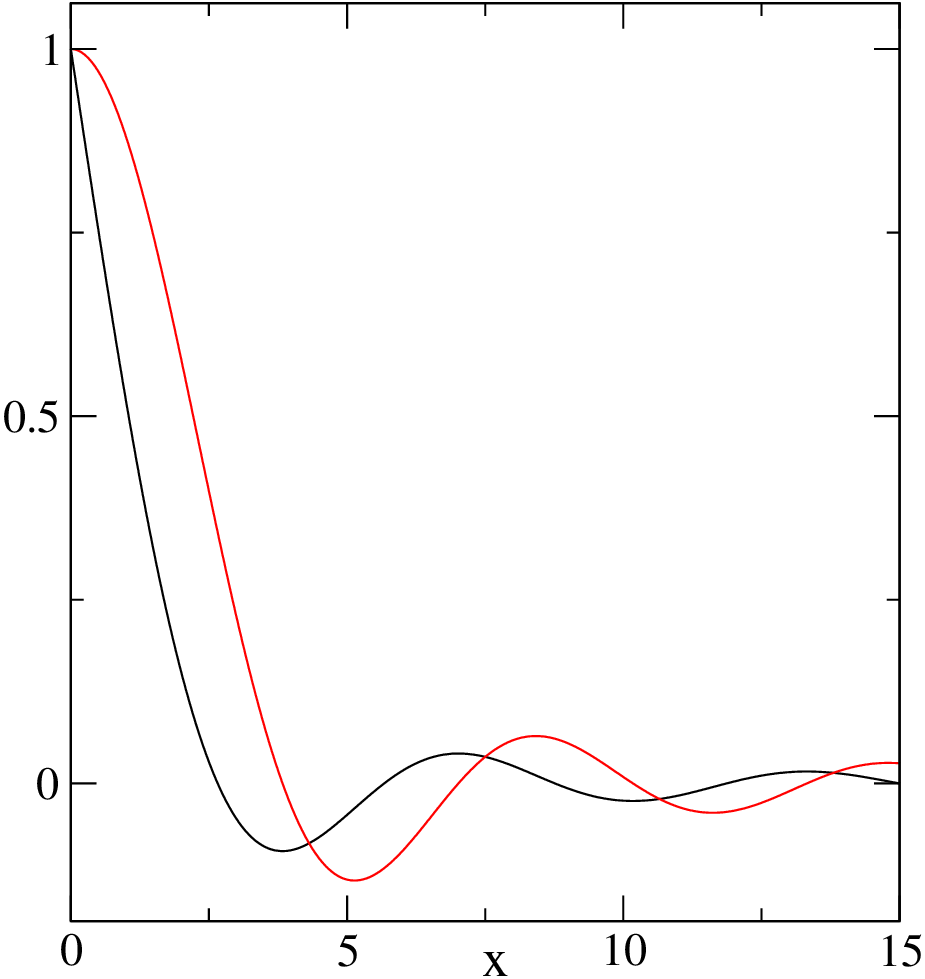}
\caption{The black line shows
(\ref{Ffunc}), whereas 
the red one depicts $\rmJ_0(x)+\rmJ_2(x)$. 
Asymptotic expressions
(\ref{Fapp}) and (\ref{JJapp})
are
 on the plotted scale 
indistinguishable
from the exact results for $x>10$.
}
 \label{Bessle_fig}
\end{figure}

The short-time dynamics of  (\ref{Esm})--(\ref{Jvecsm})
can  be understood by taking a look at 
figure \ref{Bessle_fig}.
The long-time dynamics of  (\ref{Esm})--(\ref{Jvecsm})
is captured by  the following  formulas
that can be derived with the help of \cite{Ryzhik}
\begin{align}
\label{Fapp}
& F(x)= \sqrt{\frac{2}{\pi x^3}}\sin\B{x+\frac{\pi}{4}}
+O\B{x^{-5/2}},\\
& \rmJ_0(x)+\rmJ_2(x) =  2\sqrt{\frac{2}{\pi x^3}}\sin\B{x-\frac{\pi}{4}}
+O\B{x^{-5/2}}.
\label{JJapp}
\end{align}
These expressions show that (\ref{Esm})--(\ref{Jvecsm})
exhibit damped oscillations  for $t\to\infty$,
which  are somewhat
reminiscent of the dynamics seen the studies of  quantum 
transients \cite{TransientsReview}.
 The  difference between 
 the arguments of the sine functions in (\ref{Fapp}) 
 and (\ref{JJapp}) can be understood by 
 considering the dynamics of
the charge contained
 in a tiny sphere around the point $\rr=\0$.
 Its decrease (increase) is caused by the 
 outward (inward) orientation of the 
charge  current,
which the $\pi/2$ phase shift guarantees.

Finally, we compute  the total charge stored in the 
ball of radius $t^-$, which is given by 
\be
{\cal Q}(t,r<t)=
\lim_{r\to t^-}\int d\S(\rr)\cdot\EE(t,\rr)=
q\cos(mt)-q\exp(-mt)(1+mt).
\label{Qsmma}
\ee
By combining this result with  (\ref{Qfar}), we see that
\be
{\cal Q}(t,r<t) + {\cal Q}(t,r>t)={\cal Q}(t).
\ee

\subsection{Perpetual dynamics: from  escaping to  collapsing 
evolution}
\label{Perpetual_sec}

Expressions (\ref{calE}), (\ref{calJ0}), and 
(\ref{calJ}) exhibit the following 
time-reversal 
(anti)symmetry
\begin{align}
&\EE(-t,\rr)=\EE(t,\rr),\\
&{\cal J}^0(-t,r)={\cal J}^0(t,r),\\
&\JJ(-t,\rr)=-\JJ(t,\rr).
\end{align}
As a result of that, by solving 
the dynamics of 
the mean
electric field,  charge density, and
 charge current  for $t>0$,
 one actually gets the above 
 quantities also for the
 times  $t<0$
 under the following condition.
 Namely, one has to assume 
 that the Schr\"odinger 
  picture state of the system at 
  such times  is given 
  by the action of the 
  operator 
  $\exp(-\ii H t)$ onto (\ref{hpsi}),
  where $H$ is given by (\ref{Ham}).

In particular, in the context of the
specific studies discussed 
  in section \ref{Example_sec}, such 
 an extension of the dynamics to 
the  domain
$t<0$ leads to the following observations.
We could say that for
$t:-\infty\to0^-$ 
there is  a collapsing (incoming) 
regularized Coulomb field in the studied problem.
At $t=0$ the regularized 
 Coulomb field is fully formed
 and the state of the system is
 given by (\ref{hpsi}) combined with 
 (\ref{hatfL2}), (\ref{aa}), and (\ref{bb}).
 Then, for $t:0^+\to\infty$
 the dynamics of the escaping (outgoing) 
regularized Coulomb field 
is seen in the analyzed  problem.

Finally,  we note  that
the results   discussed
in this work 
   are markedly  
   distinct from
topological  \cite{MantonBook}
and 
 nontopological \cite{NugaevReview}
 soliton solutions,
 despite the fact that 
the ``charge'' profoundly impacts
both our and the 
soliton considerations.

\section{Further illustration of periodic charge 
oscillations}
\label{Further_sec}

We  discuss   here further 
concrete theoretical illustration
of 
periodic charge        
oscillations. For this purpose, 
we  briefly extend the 
studies from section 
\ref{Example_sec}
by    considering 
\be
f(\OM{k})=\B{\frac{ m}{\VAREPS{k}}}^\gamma, 
\label{hatfL}
\ee
where 
\be
\gamma=4,6,8,\cdots
\label{ggGG}
\ee
for the technical reason mentioned below
(\ref{ftfy}). 
The following comments are in order now.

First, by repeating the calculations from the beginning
of
section \ref{Example_sec}, we find that  now
\begin{align}
&1=\alpha^2 + \beta^2 \frac{e^2}{8\pi^{3/2}}\frac{\Gamma(\gamma-1)}{\Gamma(\gamma-1/2)},\\
&{\cal H}= \beta^2
m\frac{e^2}{8\pi^{3/2}}\frac{\Gamma(\gamma-3/2)}{\Gamma(\gamma-1)},
\end{align}
the amplitude of periodic charge oscillations is 
 upper bounded by
\be
\sqrt{2\pi^{3/2}\frac{\Gamma(\gamma-1/2)}{\Gamma(\gamma-1)}},
\label{qboun}
\ee
and   $\alpha$ and $\beta$ are  given by (\ref{aa}) and (\ref{bb})
with 
$\pi$   replaced by  (\ref{qboun}).

Second, the electric field potential now reads
\be
\phi(t,r; \gamma)=\frac{q}{2\pi^2r}\int_0^\infty
d\OM{k} \B{\frac{ m}{\VAREPS{k}}}^\gamma \frac{\sin(\OM{k}r)}{\OM{k}}\cos(\VAREPS{k}t).
\label{phigamma}
\ee
It turns out that it is  smoother
than
$\phi(t,r)$ studied in section \ref{Example_sec}. 
In particular, it   follows from 
the discussion in appendix \ref{Smoothness_app}
that (\ref{phigamma}) and  all its derivatives 
up to the  order $\gamma-2$ are continuous.
This implies that
(\ref{Ehpsi}) and (\ref{calE})
are equivalent,  the equalities in 
(\ref{calJ0}) and (\ref{calJ}) hold, and
the mean
electric field and  $4$-current
are continuous at $r=t$.
Despite all that,  
$\phi(t,r; \gamma)$ is non-analytic 
(e.g. it can be inferred  from  appendix 
\ref{Non_app}  
that at least some derivatives of (\ref{phigamma})
of the order  $\gamma$ are undefined  at  $r=t$).
In other words, there is a  shock wave 
 here  just as 
in section \ref{Example_sec},
a  more  gentle
one, however.

Third, we would like to mention one more 
consequence
of the above-specified  smoothness of 
$\phi(t,r; \gamma)$. Namely, as can be inferred
from (\ref{KGeq}),  $r\phi(t,r; \gamma)$ 
satisfies the $1+1$ dimensional 
Klein-Gordon
equation 
\be
\left(\frac{\partial^2}{\partial r^2} - 
\frac{\partial^2}{\partial t^2}\right)
\BB{r\phi(t,r; \gamma)}
=m^2 r\phi(t,r; \gamma),
\ee
which places the results discussed 
in this section in a 
broader context.

Fourth, we find that for 
$r\ge t>0$  
\begin{align}
\label{hatphiLexp}
&\phi(t,r; \gamma)=q\frac{\cos(m t)-P_\gamma(mr,mt)\exp(- m r)}{4\pi r},\\
&\EE(t,\rr;\gamma)=\frac{q\rhat}{4\pi r^2}\cos(mt) -
\frac{q\rhat \exp(-mr)}{4\pi r^2}(1+mr-r\partial_r)P_\gamma(mr,mt),
\label{ElargeGamma}\\
&{\cal J}^0(t,r;\gamma)=
\frac{q \exp(-mr)}{4\pi r}(m-\partial_r)^2 P_\gamma(mr,mt),
\label{J0farGamma}\\
&
\JJ(t,\rr;\gamma)=\frac{qm\rhat}{4\pi r^2}\sin(mt) +
\frac{q\rhat \exp(-mr)}{4\pi r^2}(1+mr-r\partial_r)\frac{\partial}{\partial t}P_\gamma(mr,mt),
\label{JfarGamma}
\end{align}
where $P_\gamma$ is a  finite-degree polynomial of both arguments 
(see appendix \ref{Integral_gamma_app}, 
where the  recipe   for the 
generation of these polynomials is provided 
along with some explicit results for small 
values of $\gamma$). 
We will now briefly comment upon these
results.

Expressions  
   (\ref{hatphiLexp})
and (\ref{ElargeGamma}) 
manifestly encode the 
periodic charge oscillations 
and 
it  is immediately evident
from them that  
(\ref{Qqcos})  also holds  in the discussed case.
Moreover, they 
illustrate  that the short-range 
contribution to the electric field 
potential, and so also to the
mean electric field, 
does not have to be static in the studied 
region of space.
Mean  charge density (\ref{J0farGamma}) 
is qualitatively the same as (\ref{J0far}),
both quantities are negligible in the 
long-distance limit.
 Finally,  
apart from the exponentially small
in $r$   correction,
mean charge current (\ref{JfarGamma}) 
is 
the same as (\ref{Jfar}).

\section{Relativity-related   issues}
\label{Relativity_sec}
Two features of the solutions discussed in 
sections \ref{Example_sec} and \ref{Further_sec}
may seem to be a bit
counterintuitive if  one takes into account 
that the Lorentz invariant theory is 
studied in this paper.

First, there is the  question of 
how the periodic oscillations of the long-range 
component of the mean electric field 
can simultaneously happen at arbitrarily large 
distances. Second, there is the question of 
how the charge could periodically disappear
and reappear 
as if it would be instantly transferred all the way 
to spatial infinity. We will  discuss these
issues below.

\subsection{Long-range oscillations of 
electric field}

The first terms in     (\ref{Elarge})
and (\ref{ElargeGamma})
have a curious property. Namely, they describe 
 global, i.e.
simultaneous for all $r>t$, oscillations of the Coulomb
field. At first sight, such oscillations 
may look as if they were  caused by 
the infinite propagation speed of some signal.
However, their presence  does
not contradict special 
theory of relativity because
the Coulomb field, at every point in the $r>t$ region, oscillates 
in a predefined manner.

\begin{figure}[t]
\centering 
\includegraphics[width=\pref\columnwidth,clip=true]{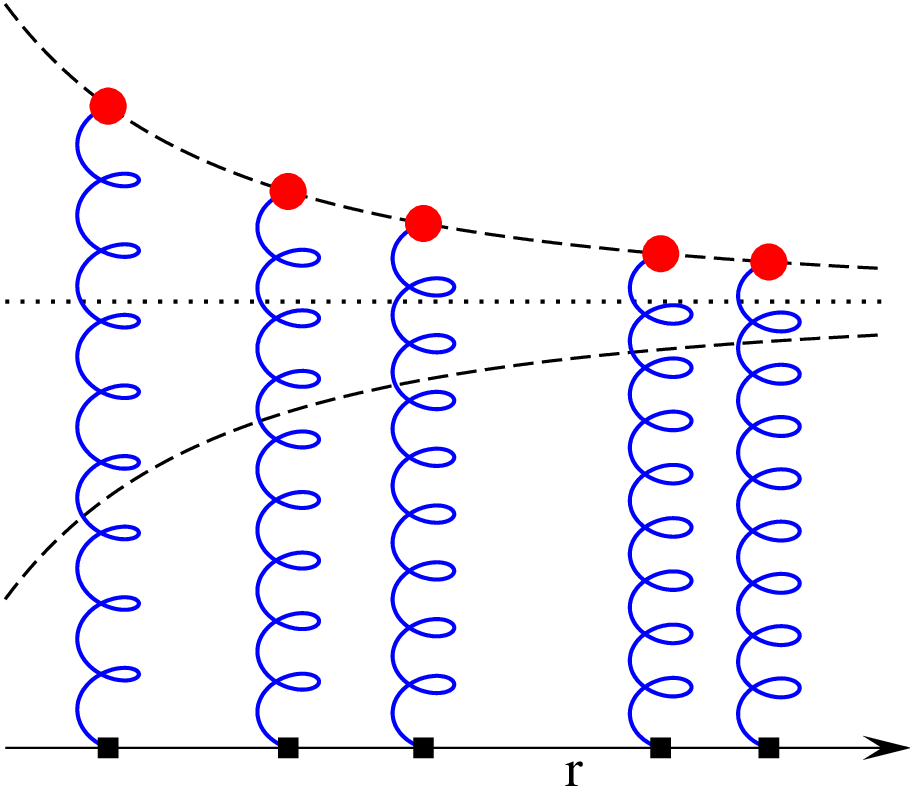}
\caption{The schematic illustration of the 
mechanical 
model   of the simultaneous 
Coulomb field oscillations. 
The springs are depicted in blue.
The  oscillating masses are presented as  red dots. 
The black dotted line goes through 
the equilibrium positions of the oscillating masses
(oscillations take place in the vertical 
direction).
The initial displacements of
the masses from their equilibrium positions
are  chosen along the upper black dashed line,
which is assumed to be described  by  $A/r^2$ ($r$ is
the position of each  spring along the
horizontal axis, $A$ is a positive constant). 
As the lower dashed line depicts  $-A/r^2$, 
the masses oscillate between 
the two dashed lines. If the angular frequency 
of these oscillations is equal to  $\omega$, then 
the   displacements of the masses 
are given by $A\cos(\omega t)/r^2$,
which is qualitatively the same as the 
formula describing
 the radial component 
 of the periodically oscillating 
 Coulomb field in 
(\ref{Elarge})
and (\ref{ElargeGamma}).
}
\label{oscillators_fig}
\end{figure}

We note that qualitatively the same effect can be observed 
in the setup composed of  harmonic oscillators, 
say identical spring-mass systems depicted in
figure \ref{oscillators_fig}.
Namely, one can place them 
along some line, take each one of them 
out of its  equilibrium position in  the direction perpendicular 
to that line, and release  them at one and 
the same time moment in their common  rest frame
(the last step 
can be always done by equipping each oscillator with a 
synchronized clock and a mechanism automatically 
starting the oscillations at the predefined time 
instant). By properly choosing 
the initial displacement 
of each oscillator, the mechanical model of 
the simultaneous Coulomb field oscillations 
can be  obtained.

\subsection{Charge transport}

For the sake of simplicity and 
definiteness, we will focus here on the 
discussion of the 
dynamics   carefully worked out in section 
\ref{Example_sec}
(the results presented  in section \ref{Further_sec}
can be similarly commented upon).
To begin, we would like to 
stress that the 
periodic variation of the total charge
has nothing to do with the discontinuity of
the mean $4$-current captured by 
(\ref{sHH1}) and (\ref{sHH2}).
In other words, the periodically
oscillating charge is not appearing or 
disappearing because it comes in and out 
of  the   ``shock portal''.  
For example,
this is evident from the discussion in 
section \ref{Further_sec}, where 
exactly the same 
periodic oscillations of the total 
charge  happen in the presence 
of the continuous  mean $4$-current.

Having said all that, we turn our attention to 
mean charge current (\ref{Jfar}),
computing its 
flux   through  the sphere of the 
radius $R>t$
\be
\int d\S(\boldsymbol{R})\cdot\JJ(t,\boldsymbol{R})= 
q  m\sin( m t).
\label{SSjj}
\ee
Since this  result 
is independent of  $R$, it  suggests
that the periodically oscillating 
charge 
outflows (inflows) to (from) spatial infinity 
without being accumulated anywhere in the 
$r>t$ region. As this happens, 
the  charge 
inside the $r<t$ region accordingly 
decreases (increases), which is evident 
from (\ref{Qsmma}).

\begin{figure}[t]
\centering 
\includegraphics[width=\pref\columnwidth,clip=true]{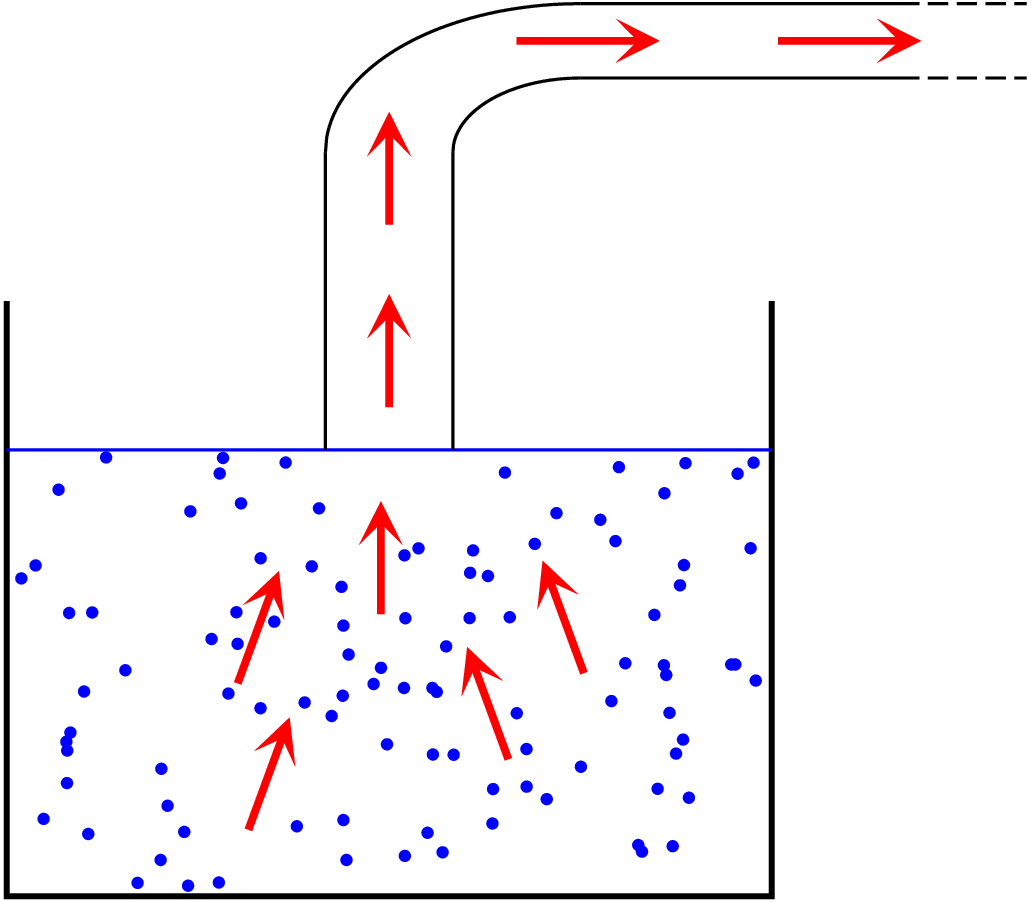}
\caption{The schematic hydrodynamic 
illustration of the empty hose paradox.
The container is filled with
the  fluid depicted by blue dots.
 The fluid either leaves or 
 enters the container via 
the hose.
Such a situation is
paradoxical because the   density
of the fluid, unlike its velocity
field depicted by red arrows, 
vanishes   inside the hose.
In the Proca theory context, 
the container represents the $r<t$ region,
the hose represents  the $r>t$
region, and the fluid stands for 
the periodically oscillating charge.
}
 \label{empty_paradox}
\end{figure}

However, the claim that  the 
periodically oscillating charge
flows through the region $r>t$ could
be confusing because according to 
(\ref{J0far})
such a charge 
does not seem to 
be ever present there.\footnote{Charge      density  (\ref{J0far})
      comes from the
     time-independent Debye component of mean electric field (\ref{Elarge}),
     which has nothing to do with periodically
        oscillating flux  (\ref{SSjj}).
     The time-dependent
     Coulomb component of mean electric
     field (\ref{Elarge}) makes (\ref{SSjj})
     nonzero, but it
      does {\it not} contribute to charge density (\ref{J0far}).
}
This remark suggests that we have 
some sort of a paradox here, 
the empty hose paradox so to speak,
whose  illustration is 
presented in figure \ref{empty_paradox}.
It is our opinion that its
explanation may be the following.

To begin, we note that  
${\cal J}^0$
can be locally either positive or negative
and consider at some time  $t>0$   two spheres having
radiuses $R, R+dR$, where $R>t$ and  $dR$ is infinitesimally
small. We  envision the following process.
The charge outflowing via the larger sphere  between 
the times $t$ and $t+dt$
leaves behind the spheres the charge $-q\sin( m t) m dt$, which is 
simultaneously  ``neutralized'' by the charge 
$q\sin( m t) m dt$
flowing into the area
between the spheres via the smaller sphere. 
In other words, the very fact that the charge density can be either
positive or negative creates the opportunity that at any time moment 
the mean charge between the spheres can be  equal to
zero.
By considering the series of 
spheres with progressively larger radiuses, $t+dR$, $t+2dR$, 
$t+3dR$, 
$\cdots$, 
we can see how the charge $q\sin( m t) m dt$ may disappear from the system
between time instants  $t$ and $t+dt$.

\section{Summary}
\label{Summary_sec}

We have studied  the dynamics of 
normalizable finite-energy 
charged states 
in the Proca theory.
Two characteristic features
have been 
observed in the course of these investigations.

First, we have found 
that at large  distances the 
mean electric field in the studied 
states 
is given by  the Coulomb field
that  is periodically oscillating 
in the time domain. On the one hand, 
such a result explains the 
periodic charge oscillations.
On the other hand,  it leads  
to
the relativity-related  questions 
that we have discussed.

Second, we have found that 
 there is a genuine
shock wave propagating in the studied 
 states. Its presence breaks analyticity 
 of the mean electric field and the related 
 quantities such as the mean charge
 density and current. 

We hope  that these and other 
 results that we have presented 
 provide solid evidence 
 that the studies 
of the dynamics of 
charged states in the Proca theory constitute 
a well-defined 
mathematical  problem,    
where elegant analytical   results 
can be obtained.

If we now turn our attention to the question of
whether the discussed results can be experimentally 
approached, the situation is problematic  
for the following two reasons.

First, the causality considerations 
 exclude the possibility 
of the experimental preparation  of 
charged states.
Still, one could presumably consider 
the perpetual evolution scenario that we
have proposed,
in which the charged states are eternally 
evolving   across the Universe
as opposed to being  created in the 
laboratory.

Second, there is the question of what stable
particle 
could represent the massive vector boson described 
by the Proca theory. In this respect,
one may  consider the possibility that 
photons may have a non-zero mass
(this issue  is related to the
experimental 
studies that have been carried out 
for over two centuries so far
\cite{Gillies2004,Nieto_RMP2010}).
In fact, there exist 
various upper bounds on the photon 
mass ranging e.g. 
from $10^{-49}$kg to $10^{-54}$kg \cite{Nieto_RMP2010}.
Upon substitution of such values into (\ref{PerioD}), 
one gets an intriguing lower bound on the oscillation period
of the Coulomb component 
of the mean electric field  
in charged  states: 
    $7\times10^{-2}$s to $7\times10^3$s. 
If such a field would exist in Nature,
it would  accelerate 
 electrons, protons, ions, etc.
Thereby, it would lead to 
the generation of 
the electromagnetic radiation
(transverse photons), 
whose frequency would encode 
the  mass $m$
of the vector boson.

All in all, 
we see this work as being mainly of 
theoretical interest at the moment.
It is our opinion that
further research along these lines 
could  lead to the 
in-depth understanding 
of the charged sectors of  Proca and similar 
theories possessing a
non-trivial infrared structure.

\acknowledgments

I am indebted to  Adolfo del Campo, 
Bogdan S. Damski, and Mateusz {\L}\k{a}cki
for stimulating  comments 
about this work.
These studies have
been  supported by the Polish National
Science Centre (NCN) Grant No. 2019/35/B/ST2/00034.
The research for this publication has been also supported by
a grant from the Priority Research Area DigiWorld under the
Strategic Programme Excellence Initiative at Jagiellonian University.

\appendix

\section{Conventions}
\label{Conventions_app}

We use  the metric tensor  $\text{diag}(+---)$.
Greek and Latin indices of tensors  take values $0,1,2,3$ and   $1,2,3$,
respectively. 
 $3$-vectors are written in bold, e.g. $x=(x^\mu)=(x^0,\x)$.

We adopt the Heaviside-Lorentz system of units
and set  $\hbar=c=1$. 
Hyperbolic $\sin$, $\cos$, $\tan$, and $\cot$ are denoted as
$\sinh$, $\cosh$, $\tanh$, and $\coth$, respectively.
   $\partial_t=\partial/\partial t$, 
    $\partial_r=\partial/\partial r$, 
$\alpha^+$ ($\alpha^-$) denotes
the quantity that is infinitesimally larger (smaller)
than $\alpha$, and 
the  hermitian conjugation is represented 
by  $\hc$

\section{Integral from expression (\ref{hatphiL2})
}
\label{Integral_app}

We evaluate here 
\be
I(a,b)=
\int_0^\infty dx
\frac{\sin[a\sinh(x)]\cos[b\cosh(x)] }{\sinh(x) \cosh(x)}
\label{Iabnew}
\ee
representing the integral from (\ref{hatphiL2}) after the following change of
variables
\begin{subequations}
\begin{align}
\label{mappingA}
&\OM{k}= m\sinh(x),\\ 
&  m r= a, \  m t= b.
\label{mappingB}
\end{align}
\label{mapping}%
\end{subequations}
We will assume  below that $a>0$, $b\ge0$.

To proceed, we will be computing
\be
I_\pm(C)=
\int_C dz \frac{\exp[\ii a\sinh(z)\pm
\ii b\cosh(z)]}{\sinh(z)\cosh(z)}
\label{IZ}
\ee
over the contours  depicted in figures \ref{1komtur}
and \ref{3komtur}. 
We have chosen their  shapes in a standard way,
taking into account  that the 
integrand has simple poles at integer and half-integer
multiples of  $\ii\pi$.

\subsection{$a\ge b>0$}
\label{sub1_app}
The goal here is to prove that for $a\ge b>0$
\be
I(a,b)=\frac{\pi}{2}\BB{\cos(b)-\exp(-a)}.
\label{Iabnew1}%
\ee

\begin{figure}[t]
\centering 
\includegraphics[width=\pref\columnwidth,clip=true]{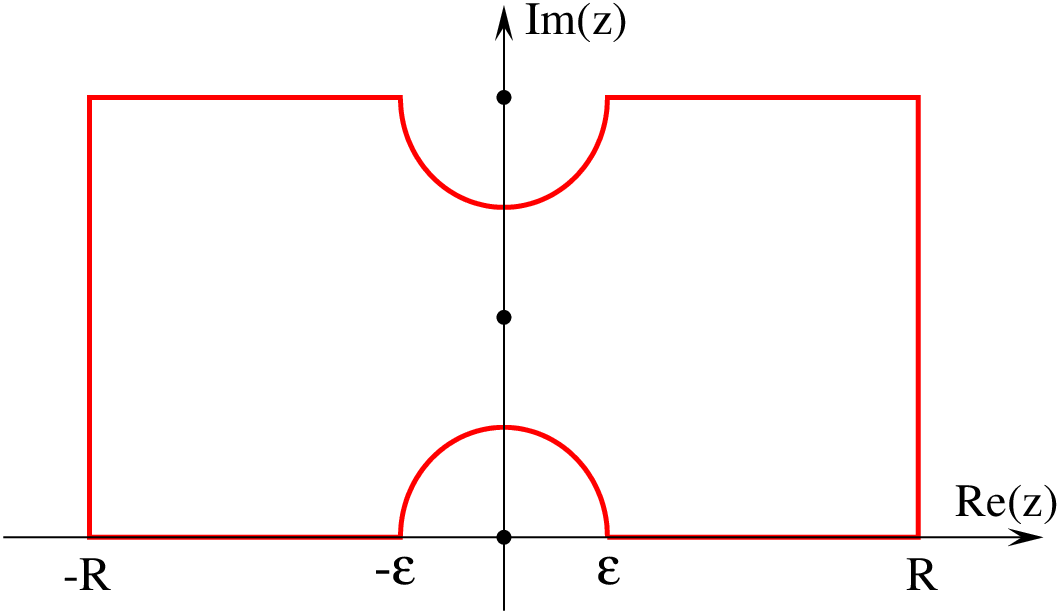}
\caption{Integration contour. 
Its horizontal segments are at $\im(z)=0,\pi$.
Curved segments are semi-circular with radius $\varepsilon$. The 
poles of the integrand are depicted by the dots
(they are located at  $z=0,\ii\pi/2,\ii\pi$).
}
\label{1komtur}
\end{figure}

The integral $I_+(C)$  is done over the contour from figure \ref{1komtur}. 
Proceeding in the standard way, we get
\be
\dashint dx \frac{\sin[a\sinh(x)+b\cosh(x)]}{\sinh(x)\cosh(x)}=
\pi\cos(b)
-\pi \exp(-a),
\label{I1pv}
\ee
where the limits of
$R\to\infty$ and $\varepsilon\to0^+$ have been taken
and
$\dashint$ stands for the Cauchy principal value 
\be
\dashint = \lim_{\varepsilon\to0^+}\B{
\int_{-\infty}^{-\varepsilon}
+\int^{\infty}_{\varepsilon}
}.
\label{CPV}
\ee
The left-hand side of (\ref{I1pv}) comes from 
integration over horizontal segments. 
The right-hand side of  (\ref{I1pv}), 
besides  the $-\pi \exp(-a)$ pole  contribution, 
comes  from
integration over semi-circular 
segments.

The vertical segments do not   contribute to (\ref{I1pv}) for the following
reasons. At $z=-R+\ii y$,
\be
\label{chsh}
\BBB{\frac{1}{\sinh(z)\cosh(z)}}=O(\exp(-2R)),
\ee

\begin{subequations}
\begin{align}
&\BBB{\exp(\ii a\sinh(z)+\ii b\cosh(z))}=
 \exp\B{-\frac{a-b}{2}\sin(y)\exp(R) +\epsilon(a,b)},\\
&
\epsilon(a,b)=-\frac{a+b}{2}\sin(y)\exp(-R)=O(\exp(-R)),
\end{align}
\label{mod1}%
\end{subequations}
and so the integrand  vanishes  in the $R\to\infty$
limit for any fixed 
\be
0\le y \le \pi
\label{yy}
\ee
being of interest here. Note that it is so
because the  $a\ge b$ case is considered.
The same conclusion applies to the vertical segment parameterized by  $z=R+\ii y$ because 
(\ref{chsh}) still holds, 
\be
\BBB{\exp(\ii a\sinh(z)+\ii b\cosh(z))   }=
 \exp\B{-\frac{a+b}{2}\sin(y)\exp(R) + \epsilon(a,-b)},
\label{mod2}
\ee
and   $a+b>0$.

The  integral $I_-(C)$  
is also done over the contour from figure \ref{1komtur}. In 
very much the same way as above, it leads to 
\be
\dashint dx \frac{\sin[a\sinh(x)-b\cosh(x)]}{\sinh(x)\cosh(x)}
=\\ \pi\cos(b)-\pi\exp(-a),
\label{I2pv}
\ee
which  immediately follows from 
the $x\to-x$ change of the integration variable in (\ref{I1pv}).
By combining  (\ref{I1pv}) and  (\ref{I2pv}), (\ref{Iabnew1}) 
can be established.

\subsection{$b>a>0$}
\label{sub2_app}
It will be first argued here that for $b>a>0$
\be
I(a,b)=\frac{\pi}{2}\sinh(a) - \int_0^{\pi/2}dx\frac{\sinh[a\cos(x)]\sin[b\sin(x)]}{\sin(x)\cos(x)}
\label{Iab_1}
\ee
and then it will be shown that
\be
I(a,b)=\frac{\pi}{2}[\cos(a)-\exp(-a)] \\ - 
\frac{\pi}{2} \int_a^b dy \int_0^a dx \rmJ_0\B{\sqrt{y^2-x^2}}.
\label{III2I}
\ee
These equations  trigger the following  
comment.

\begin{figure}[t]
\centering 
\includegraphics[width=\pref\columnwidth,clip=true]{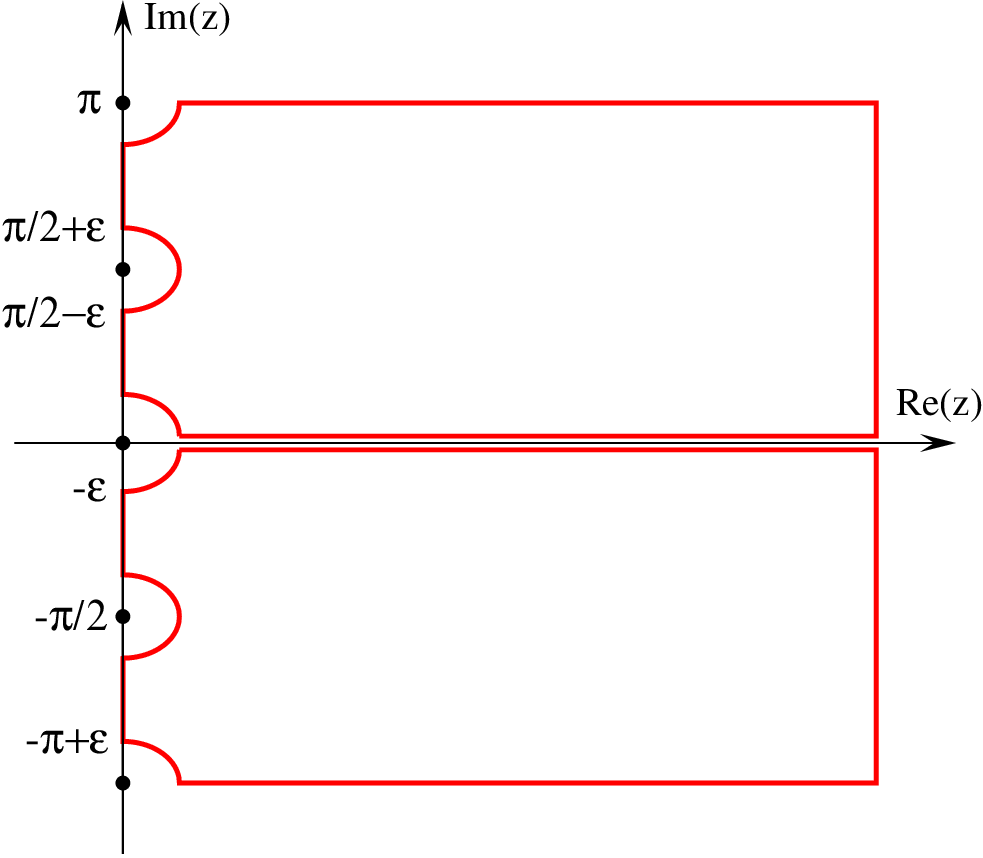}
\caption{Two integration contours. 
Their horizontal and vertical segments are at $\im(z)=-\pi,0,\pi$ 
and $\re(z)=0,R$, respectively.
Curved segments, half- and  quarter-circular, have radius $\varepsilon$. They 
go around the poles of the integrand, which are depicted by black dots.
The lower contour is the mirror image of the upper one.
}
\label{3komtur}
\end{figure}

We  are unaware of the closed-form 
expressions for the integrals in (\ref{Iab_1})
and (\ref{III2I}). Nonetheless,
these formulas are arguably more useful than (\ref{Iabnew}).
It is so because  their derivatives 
with respect to 
 $a$ and   $b$
can be easily computed  and they are 
manifestly finite. This  is
particularly clearly seen from  (\ref{Iab_1}), 
where it is  evident that the 
integrand and all its derivatives 
with respect to    $a$ and 
$b$ are defined and continuous\footnote{It should       be understood that 
      $\sin[b\sin(x)]/\sin(x)=b(1-b^2\sin^2(x)/3!+\cdots)$
           near and at $x=0$.
      Similarly,
      $\sinh[a\cos(x)]/\cos(x)=a(1+a^2\cos^2(x)/3!+\cdots$)
      near and at $x=\pi/2$.},
which in the case of the {\it proper} integral 
is basically all one needs to know to be able to 
take the derivatives 
under the integral symbol.\footnote{The
issue of interchanging derivatives
 and integrals is discussed in numerous
 textbooks (see e.g. \cite{LangBook}).
The particularly comprehensive  presentation
of this topic can be found in Polish in  \cite{JarnickiWyklad_art}.}
By contrast, the differentiation of
(\ref{Iabnew}) is 
more complicated because we deal in 
such a case with an  {\it improper} integral 
over an  unbounded  interval.

We start the discussion of 
 the derivation of 
(\ref{Iab_1})
with the computation of $I_+(C)$ over the upper contour in figure
\ref{3komtur}. Note that 
the contour from figure \ref{1komtur} cannot 
be now used for the evaluation of $I_+(C)$
 because the integrand on its 
 vertical segment at $\re(z)=-R$ 
quickly grows in the $R\to\infty$ limit when $b>a$,
which is seen from (\ref{mod1}).
 
We  proceed  similarly as in section \ref{sub1_app}. Introducing 
\be
\Upsilon_\pm(\varepsilon)=
\int_\varepsilon^\infty dx\frac{\sin[a\sinh(x)\pm b\cosh(x)]}{\sinh(x)\cosh(x)}
 \pm \frac{\ii}{2}
\B{\int_\varepsilon^{\pi/2-\varepsilon}+\int_{\pi/2+\varepsilon}^{\pi-\varepsilon}}
 dx\frac{\exp[\mp a\sin(x)+\ii
b\cos(x)]}{\sin(x)\cos(x)},
\label{Upsilonn}
\ee
we  arrive at 
\be
\lim_{\varepsilon\to0^+}\Upsilon_+(\varepsilon)=\frac{\pi}{2}\BB{\cos(b)-\exp(-a)}.
\label{qeeq1} 
\ee
The first term in  (\ref{Upsilonn}) 
comes from integration
over horizontal segments of the upper contour in figure \ref{3komtur}. 
The rest  of (\ref{Upsilonn}) comes from 
 integration  
over the vertical segments at $\re(z)=0$. The right-hand side of (\ref{qeeq1})
comes from integration over semi- and quarter-circular segments. The vertical segment at 
$\re(z)=R$ does not contribute, which is seen from  (\ref{mod2}). 
The limit of $R\to\infty$ has been  taken in the above expressions.

Next, we integrate  $I_-(C)$  over the lower contour in figure
\ref{3komtur}. The key observation here is that integration over the vertical 
segment at $\re(z)=R$ does not contribute. It is so because 
for $z=R-\ii y$, where again (\ref{yy}) is used, we have (\ref{chsh}) and 
\be
\BBB{\exp(\ii a\sinh(z)-\ii b\cosh(z)) }=
\exp\B{-\frac{b-a}{2}\sin(y)\exp(R) - \epsilon(a,b)}.
\ee
Thereby  the integrand   vanishes in the $R\to\infty$ limit for $b>a$
considered now.
Proceeding similarly as above, we arrive at  
\be
\lim_{\varepsilon\to0^+}\Upsilon_-(\varepsilon)=\frac{\pi}{2}\BB{\exp(a) - \cos(b)}.
\label{qeeq2} 
\ee
By combining (\ref{qeeq1}) and (\ref{qeeq2}), 
(\ref{Iab_1})
can be  established.

Result (\ref{Iab_1}) can be used for expressing
the  integral of interest 
in terms of the Bessel function. Namely, 
denoting by $\tilde I(a,b)$ 
the integral in  (\ref{Iab_1}),  we find that  
\be
\frac{\partial^2}{\partial a\partial b}\tilde{I}(a,b)=\frac{\pi}{2}
\rmJ_0\B{\sqrt{b^2-a^2}}
\label{dadb}
\ee
after taking into account  formula 3.996.3 of 
\cite{Ryzhik} 
\be
\int_0^{\pi/2} dx
\cosh[a\cos(x)]\cos[b\sin(x)]=\frac{\pi}{2}\rmJ_0\B{\sqrt{b^2-a^2}}.
\label{intJ0}
\ee
Solving (\ref{dadb}) with the  boundary conditions 
$\lim_{a\to0}\tilde{I}(a,b)=\lim_{b\to0}\tilde{I}(a,b)=0$, we
get
\be
\tilde{I}(a,b)=\frac{\pi}{2}\int_0^a dx \int_0^b dy \rmJ_0\B{\sqrt{y^2-x^2}},
\ee
where it has to be understood that for $y<x$ we have
$\rmJ_0(\sqrt{y^2-x^2})=\rmI_0(\sqrt{x^2-y^2})$
with 
$\rmI_n$ representing the modified 
Bessel function of the first kind of order $n$.
  Simple manipulations, involving the following variant of (\ref{J0sin})
\be
\int_0^x dy \rmI_0\B{\sqrt{x^2-y^2}} = \sinh(x),
\ee
lead to (\ref{III2I}). 
Another  potentially useful 
representation of (\ref{Iab_1})
is given by (\ref{NewRepJ}).

\subsection{$a>b=0$}
One may easily verify with the 
help of appendix \ref{sub1_app} that 
\be
I(a,0)= \frac{\pi}{2}\BB{1-\exp(-a)} \for a>0.
\label{IabZERO}
\ee

\subsection{From continuity to discontinuity across $a=b$}
\label{Disc_app}
With the help of (\ref{Iabnew1}) and (\ref{III2I}), we find
that 
\begin{align}
\label{IabCont0}
& \lim_{a\to b^-} I(a,b)= \lim_{a\to b^+} I(a,b),\\
\label{IabConta}
& \lim_{a\to b^-} \frac{\partial}{\partial a}I(a,b)
= \lim_{a\to b^+} \frac{\partial}{\partial a}I(a,b),\\
\label{IabContb}
& \lim_{a\to b^-} \frac{\partial}{\partial b}I(a,b)
=\lim_{a\to b^+} \frac{\partial}{\partial b}I(a,b),\\
\label{IabContaa}
& \lim_{a\to b^-} \frac{\partial^2}{\partial a^2}I(a,b)
= \frac{\pi}{2}+ \lim_{a\to b^+} \frac{\partial^2}{\partial a^2}I(a,b),\\
\label{IabContab}
& \lim_{a\to b^-} \frac{\partial^2}{\partial a\partial b}I(a,b)
= -\frac{\pi}{2} + \lim_{a\to b^+} \frac{\partial^2}{\partial a\partial b}I(a,b),\\
\label{IabContbb}
& \lim_{a\to b^-} \frac{\partial^2}{\partial b^2}I(a,b)
= \frac{\pi}{2} + \lim_{a\to b^+} \frac{\partial^2}{\partial b^2}I(a,b).
\end{align}

\section{Integral from expression (\ref{hatphiL2}): differentiation 
under the integral sign}
\label{Sign_app}

We introduce  the following notation
\be
I_{\alpha\beta}(a,b)=\int_0^\infty dx 
\frac{\partial^\alpha}{\partial a^\alpha}
\frac{\partial^\beta}{\partial b^\beta}
\frac{\sin[a\sinh(x)]\cos[b\cosh(x)] }{\sinh(x)\cosh(x)},
\label{IALBE}
\ee
where $\alpha,\beta\in\mathbbm{N}_0$ and $I_{\alpha\beta}$
should not be confused with
the modified
Bessel function of the first kind 
$\rmI_n$. The
goal here is to compute 
$I_{\alpha\beta}(a,b)$ for $1\le\alpha+\beta\le2$
 and to compare it 
to  
\be
\partial_a^\alpha\partial_b^\beta I(a,b)
=\frac{\partial^\alpha}{\partial a^\alpha}
\frac{\partial^\beta}{\partial b^\beta}
\int_0^\infty dx 
\frac{\sin[a\sinh(x)]\cos[b\cosh(x)] }{\sinh(x)\cosh(x)},
\ee
which can be obtained from the results presented
in appendix \ref{Integral_app}. This will be 
done under the tacit assumption that $a,b>0$.

Whenever the  order of
differentiation  and integration 
does 
not matter, i.e.  when 
\be
\BB{\frac{\partial^\alpha}{\partial a^\alpha}
\frac{\partial^\beta}{\partial b^\beta},
\int_0^\infty dx}=0
\label{[]=0}
\ee
in the studied problem, 
we will have 
$I_{\alpha\beta}(a,b)=\partial_a^\alpha\partial_b^\beta I(a,b)$.
The  formal
determination of  when it happens,
which would reduce the number of 
integrals evaluated in this section, 
is not so 
trivial because we deal
here with  improper
integrals over an unbounded 
interval.\footnote{The term formal refers here to 
      ``theorem-based'' considerations,
      where no explicit
      integral evaluations are carried out.
      The considerations of such a type can be
      found in appendix \ref{Smoothness_app},
      but as they stand they are not sophisticated enough
      to yield useful information in the context of
      the studies discussed in appendix \ref{Sign_app}.} 
Therefore, as far as we see it, 
 the easiest way to
 proceed 
here is to  do 
a  direct, i.e. brute force, 
evaluation of 
$I_{\alpha\beta}(a,b)$.
The comparison of such obtained results to 
$\partial_a^\alpha\partial_b^\beta I(a,b)$
could be then  straightforwardly
done [see e.g.
the comment  below (\ref{III2I})].

The  results reported below come from the evaluations
of  
\be
\int_C 
dz f(z)\exp[\ii a\sinh(z)\pm
\ii b\cosh(z)],
\label{IZPM}
\ee
which proceed similarly as the 
carefully discussed calculations 
reported in appendices \ref{sub1_app} and \ref{sub2_app}.
Thus, we will not dwell on the technical details
of such computations.
We will, however,
briefly comment upon two issues.

First, (\ref{IZPM}) will be evaluated 
over the   contours from figures \ref{1komtur}
and \ref{3komtur} subjected to small alterations
(replacements of half- and quarter-circular 
segments by line segments in the absence of
poles).
The contour from figure \ref{1komtur}
(figure \ref{3komtur}) 
will be re-used
for   $a\ge b$ ($b>a$).

Second,  $f(z)$  will be given by 
$1/\cosh(z)$, $1/\sinh(z)$, 
$\tanh(z)$, $1$, and $\coth(z)$ during 
the evaluation of (\ref{IAa}),
(\ref{IBb}), (\ref{IAAaa}), 
(\ref{IABab}), and (\ref{IBBbb}),
respectively.
This means that  
in the last three cases 
 $|f(\pm R+\ii y)|\to1$  
 in the   $R\to\infty$ 
 limit when $0\le y \le \pi$.
This is of crucial 
importance for the case of  $a=b$. Namely,
it implies that 
for 
$a=b$  
the vertical segment of the contour 
at $\re(z)=-R$  ($\re(z)=R$)
 will
contribute to  integral (\ref{IZPM}) when the $+$ ($-$)
sign will be chosen. 
This fact stands behind the $\pi/4$ shift in 
(\ref{IaaEv2}), (\ref{IabEv2}), and (\ref{IbbEv2}).

\subsection{First order derivatives}
\label{First_app}
We have found that 
\be
I_{10}(a,b)=\int_0^\infty dx \frac{\cos
[a\sinh(x)]\cos[b\cosh(x)]}{\cosh(x)}
\label{IAa}
\ee
is equal to
\begin{subequations}
\begin{align}
\label{IaEv1}
& \frac{\pi}{2}\exp(-a)    \for a\ge b,\\
& \frac{\pi}{2}\exp(-a) -\frac{\pi}{2}\int_a^b dx \rmJ_0\B{\sqrt{x^2-a^2}}   \for b>a.
\end{align}
\label{IaEv}%
\end{subequations}

Moreover, we have found that 
\be
\label{IBb}
I_{01}(a,b)=-\int_0^\infty dx  \frac{\sin[a\sinh(x)]\sin[b\cosh(x)]}{\sinh(x)}
\ee
is equal to
\begin{subequations}
\begin{align}
\label{IbEv1}
&  -\frac{\pi}{2}\sin(b)  \for a\ge b,\\
&   -\frac{\pi}{2}\sin(b) +\frac{\pi}{2}\int_a^b dx \rmJ_0\B{\sqrt{b^2-x^2}}  \for b>a.
\end{align}
\label{IbEv}%
\end{subequations}

As can be easily checked, (\ref{IaEv}) and (\ref{IbEv})
agree everywhere  with $\partial_aI(a,b)$ and 
$\partial_bI(a,b)$, respectively.
Note that we refer here to $\partial_aI(a,b)$ and
$\partial_bI(a,b)$ obtained via 
differentiation of 
(\ref{Iabnew1}) and either (\ref{Iab_1}) or 
(\ref{III2I}).
We mention in passing that 
despite the piecewise 
results for    $I(a,b)$ presented in
appendices \ref{sub1_app} and \ref{sub2_app},
\be
\partial_a I(b,b)=\lim_{\epsilon\to0}
\frac{I(b+\epsilon,b)-I(b,b)}{\epsilon}
\ee
and
\be
\partial_b I(b,b)=\lim_{\epsilon\to0}
\frac{I(b,b+\epsilon)-I(b,b)}{\epsilon}
\ee
are well-defined.
As there are no surprises here, 
we shall 
not discuss these findings any further.

\subsection{Second order derivatives}
\label{Sign_app2}
\label{Iaa_app}
We have found that 
\be
I_{20}(a,b)=-\int_0^\infty dx \tanh(x) \sin[a\sinh(x)]\cos[b\cosh(x)]
\label{IAAaa}
\ee
is equal to
\begin{subequations}
\begin{align}
\label{IaaEv1}
& -\frac{\pi}{2}\exp(-a)    \for a>b,\\
\label{IaaEv2}
& -\frac{\pi}{2}\exp(-a)+\frac{\pi}{4}    \for a=b,\\
\label{IaaEv3}
& -\frac{\pi}{2}\exp(-a)+\frac{\pi}{2}
 -\frac{a\pi}{2}\int_a^b dx \frac{\rmJ_1\B{\sqrt{x^2-a^2}}}{\sqrt{x^2-a^2}}       \for b>a.
\end{align}
\label{IaaEv}%
\end{subequations}

Then, we have found that 
\be
\label{IABab}
I_{11}(a,b)=-\int_0^\infty dx  \cos[a\sinh(x)]\sin[b\cosh(x)]
\ee
is equal to
\begin{subequations}
\begin{align}
\label{IabEv1}
&  0  \for a>b,\\
\label{IabEv2}
&  -\frac{\pi}{4}  \for a=b,\\
&  -\frac{\pi}{2}\rmJ_0\B{\sqrt{b^2-a^2}}  \for b>a.
\end{align}
\label{IabEv}%
\end{subequations}

Finally, we have found that
\be
I_{02}(a,b)=-\int_0^\infty dx \coth(x) \sin[a\sinh(x)]\cos[b\cosh(x)]
\label{IBBbb}
\ee
is equal to
\begin{subequations}
\begin{align}
\label{IbbEv1}
&-\frac{\pi}{2}\cos(b)  \for a>b,\\
\label{IbbEv2}
&-\frac{\pi}{2}\cos(b)+\frac{\pi}{4} \for a=b,\\
\label{IbbEv3}
&-\frac{\pi}{2}\cos(b) +\frac{\pi}{2} 
-\frac{b\pi}{2}\int_a^b dx \frac{\rmJ_1\B{\sqrt{b^2-x^2}}}{\sqrt{b^2-x^2}}
\for b>a.
\end{align}
\label{IbbEv}%
\end{subequations}

Simple calculations show that 
everywhere except at $a=b$
 (\ref{IaaEv}), (\ref{IabEv}), 
 and (\ref{IbbEv}) are equal to 
$\partial^2_aI(a,b)$, 
$\partial_a\partial_bI(a,b)$, 
and $\partial^2_bI(a,b)$, respectively.
Note that we refer here to 
$\partial^2_aI(a,b)$, 
$\partial_a\partial_bI(a,b)$, and 
$\partial^2_bI(a,b)$
obtained via
differentiation of
(\ref{Iabnew1}) and
either (\ref{Iab_1}) or 
(\ref{III2I}).
At $a=b$, we find that $\partial^2_aI(a,b)$,
$\partial_a\partial_bI(a,b)$,
and $\partial^2_bI(a,b)$ are
undefined,
which is in  stark contrast to what we
have in (\ref{IaaEv2}),
(\ref{IabEv2}), and 
(\ref{IbbEv2}).
This is seen from the fact that
the limits 
\begin{align}
\label{11}
&\lim_{\epsilon\to0}
\frac{\partial_a I(b+\epsilon,b)-\partial_aI(b,b)}{\epsilon},\\
&\lim_{\epsilon\to0}
\frac{\partial_b I(b+\epsilon,b)-\partial_b I(b,b)}{\epsilon},\\
&\lim_{\epsilon\to0}
\frac{\partial_b I(b,b+\epsilon)-\partial_b I(b,b)}{\epsilon}
\label{33}
\end{align}
do not exist.

Finally, we would like to  mention 
the following identity 
\be
I_{20}(a,b)-I_{02}(a,b)=I(a,b).
\label{KGintegrand}
\ee
After putting   (\ref{IaaEv3})
and 
(\ref{IbbEv3}) into  it, 
the new representation 
of (\ref{Iab_1}) is obtained
\be
\frac{\pi}{2}\BB{\cos(b)-\exp(-a)} 
+\frac{\pi}{2}\int_a^b dx 
\BB{b\frac{\rmJ_1\B{\sqrt{b^2-x^2}}}{\sqrt{b^2-x^2}}
-a\frac{\rmJ_1\B{\sqrt{x^2-a^2}}}{\sqrt{x^2-a^2}}}.
\label{NewRepJ}
\ee

\section{Integral from expression (\ref{phigamma}): insights
relevant for periodic charge oscillations}
\label{Integral_gamma_app}

We briefly discuss below 
\be
I(a,b;\gamma)=\int_0^\infty dx
\frac{\sin[a\sinh(x)]\cos[b\cosh(x)] }{\sinh(x) \cosh^{\gamma-1}(x)}
\label{ftfy}
\ee
representing the integral from (\ref{phigamma}) under  mapping 
(\ref{mapping}). We focus here on  the
$a\ge b>0$ case, which is 
of interest in the context of the 
discussion of periodic charge oscillations. The 
values of $\gamma$ considered here are chosen so as to 
enable  the  evaluation of 
(\ref{ftfy}) by a straightforward extension 
of the procedure discussed in appendix \ref{sub1_app}.
As can be easily checked, this is achieved when 
 $\gamma=4,6,8,\cdots$.

By repeating the calculations from section \ref{sub1_app},
integrating this time on the complex plane 
\be
\frac{\exp[\ii a\sinh(z)\pm\ii b\cosh(z)]}{\sinh(z)\cosh^{\gamma-1}(z)},
\ee
we find that for $a\ge b>0$ and the above-mentioned
values of $\gamma$
\be
I(a,b;\gamma)=\frac{\pi}{2}\cos(b)
+\frac{\pi}{2}
\text{Res}\B{\frac{\cos[b\cosh(z)]\exp[\ii a\sinh(z)]}{\sinh(z)\cosh^{\gamma-1}(z)},\frac{\ii\pi}{2}}.
\label{IAB}
\ee
It is then easy to argue 
that the residue from (\ref{IAB}) can be written
as $-P_\gamma(a,b)\exp(-a)$, where $P_\gamma(a,b)$ is a polynomial 
in $a$ and $b$. For example,
\begin{align}
&P_4(a,b)=1+\frac{a}{2}-\frac{b^2}{2},\\
&P_6(a,b)=1+\frac{a^2}{8}-\frac{b^2}{2}+\frac{b^4}{24}+\frac{5a}{8}-\frac{ab^2}{4},\\
&P_8(a,b)=1 + \frac{11 a}{16} + \frac{3 a^2}{16} + 
\frac{a^3}{48} - \frac{b^2}{2} - \frac{5 a b^2}{16} 
- \frac{a^2 b^2}{16} + \frac{b^4}{24} + \frac{a b^4}{48} - 
\frac{b^6}{720}.
\end{align}
We note that  the  highest power
of $a$ and $b$ in $P_\gamma(a,b)$ is $\gamma/2-1$  and $\gamma-2$, 
which can be easily proved.

\section{Integral from expression (\ref{phigamma}): smoothness issues }
\label{Smoothness_app}
We will discuss now 
the derivatives   of 
\be
\widehat I(a,b; \gamma) = 
\int_0^\infty
d\omega \frac{ \cos\B{b\sqrt{1+\omega^2}} }{(1+\omega^2)^{\gamma/2}}   
\frac{\sin(a \omega )}{a\omega},
\ee
where
\be
 \gamma=4,6,8,\cdots
 \label{gagg}
\ee
so as to make the following discussion
relevant to the considerations from section \ref{Further_sec}.
We remark  that 
under  mapping (\ref{mapping})
\be
{\widehat I}(a,b; \gamma)=I(a,b; \gamma)/a=\frac{2\pi^2}{q m}\phi(t,r;\gamma),
\ee
where $I(a,b; \gamma)$ and $\phi(t,r;\gamma)$ are  given by (\ref{ftfy}) and
(\ref{phigamma}), respectively.

To begin, we consider
${\cal I}:\Omega\times[0,\infty)\to\RR$, where 
\be
{\cal I}(a,b,\omega;\gamma)=
\frac{ \cos\B{b\sqrt{1+\omega^2}} }{(1+\omega^2)^{\gamma/2}}
\frac{\sin(a \omega )}{a\omega},
\ee
$\Omega\ni(a,b)$ is an open set in $\RR^2$
\be
\Omega=(0,\infty)\times(0,\infty),
\label{abinterv}
\ee
and 
\be
\frac{\sin(a \omega )}{a\omega}=1-\frac{(a\omega)^2}{3!}+\cdots
\ee
near and at $a\omega=0$.
These definitions  lead to the following observations.

First, as far as the differentiability class
of ${\cal I}$ is concerned, we observe  that 
${\cal I}(a,b,\omega;\gamma)\in C^{\gamma-2}(\Omega)$
for any $\omega\in[0,\infty)$.

Second, the map 
\be\Omega\times[0,\infty)\ni(a,b,\omega)\mapsto
\partial^\alpha_a\partial^\beta_b{\cal I}(a,b,\omega;\gamma)
\in\RR
\ee
is continuous for all 
\begin{subequations}
\begin{align}
& \alpha+\beta\le\gamma-2,\\
& \alpha,\beta\in \mathbbm{N}_0.
\end{align}
\label{ineq}%
\end{subequations}
Note that  $\partial_x^\alpha=(\partial_x)^\alpha=\partial^\alpha/\partial
x^\alpha$, where  $x=a,b$.

Third, 
$|\partial^\alpha_a\partial^\beta_b{\cal I}(a,b,\omega;\gamma)|$ 
is upper bounded, 
for all (\ref{ineq}) and $(a,b,\omega)\in\Omega\times[0,\infty)$,
by the continuous function 
$u_{\alpha+\beta}:[0,\infty)\to\RR$ such that 
$\smallint_0^\infty d\omega u_{\alpha+\beta}(\omega;\gamma)<\infty$.
Namely, 
\be
u_{\alpha+\beta}(\omega; \gamma)=\left\{
\begin{array}{lcl}
1 & \for & 0\le\omega\le1 \\
\omega^{\alpha+\beta-\gamma} & \for & \omega>1
\end{array}
\right..
\label{qq11}
\ee
Such a result has been  obtained with the help of 
the following  inequality
\be
\left|\frac{d^n}{dx^n}\frac{\sin(x)}{x}\right|\le
\frac{1}{1+n}\le1,
\label{inEQ}
\ee
which is valid for $n\in\mathbbm{N}_0$ 
and $x\in\RR$ (see section 3.4.24 of \cite{Mitrinovic}
and references therein).

These  three observations
are sufficient for concluding  that
\be
\frac{\partial^\alpha}{\partial a^\alpha}
\frac{\partial^\beta}{\partial b^\beta}
\widehat I(a,b;\gamma)= 
\int_0^\infty d\omega 
\frac{\partial^\alpha}{\partial a^\alpha}
\frac{\partial^\beta}{\partial b^\beta}
{\cal I}(a,b,\omega;\gamma)
\label{comDER}
\ee
and  
$\widehat I(a,b;\gamma)\in C^{\gamma-2}(\Omega)$
\cite{JarnickiWyklad_art}. 
Both conclusions are stated 
under the assumptions that (\ref{ineq})
holds 
and $(a,b)\in\Omega$. 
In the following, we provide some
 comments related to (\ref{comDER}).

\subsection{Non-analyticity}
\label{Non_app}
We observe  that (\ref{gagg}) and 
(\ref{comDER}) allow us to write
\be
\partial_b^{\gamma-2} \widehat I(a,b;\gamma)= 
(-1)^{\gamma/2-1}I(a,b)/a
\label{d2Igamma=I}
\ee
for $(a,b)\in\Omega$.  
Taking into account 
(\ref{IabContaa})--(\ref{IabContbb}) and 
the nonexistence of limits (\ref{11})--(\ref{33}), 
it is easy to argue via  (\ref{d2Igamma=I})
that 
 $\widehat I(a,b;\gamma)$ is non-analytic.
We mention in passing that 
 the fact that the considered 
values of $\gamma$ are even is crucial for the 
establishment of (\ref{d2Igamma=I}).

\subsection{Klein-Gordon equation}
\label{KG_app}
We note that (\ref{comDER}) can be used 
for showing  
that $I(a,b;\gamma)$ satisfies  the  
$1+1$ dimensional 
Klein-Gordon equation 
\be
\frac{\partial^2}{\partial a^2}I(a,b;\gamma)-\frac{\partial^2}{\partial b^2}I(a,b;\gamma)
=I(a,b;\gamma),
\label{KGeq}
\ee
where $a$ plays the role of the spatial coordinate, 
$b$ plays the role of time, 
 the mass is set to unity, $(a,b)\in\Omega$, 
 and  (\ref{gagg})  is assumed.
For $\gamma=2$, $I(a,b;\gamma)$ turns into $I(a,b)$,
which satisfies the Klein-Gordon equation for all 
$(a,b)\in\Omega$ except
$a =  b$ 
(this can be inferred from the
discussion in appendix \ref{Sign_app}).
 There is, however, the
weaker form of the Klein-Gordon
equation that holds 
without such a restriction when $\gamma=2$
 (\ref{KGintegrand}).


%

\end{document}